\documentclass[11pt]{article}
\title{Non-Existence of Time-Periodic Solutions of the Dirac Equation
in an Axisymmetric Black Hole Geometry}
\author{F.\ Finster, N.\ Kamran\thanks{Research supported by NSERC grant
\# RGPIN 105490-1998.},
J.\ Smoller\thanks{Research supported in part by the NSF, Grant No.\ 
DMS-G-9802370.}, and S.-T.\ Yau\thanks{Research supported in part 
by the NSF, Grant No.\ 33-585-7510-2-30.}
\date{May 1999}}
\newtheorem{Def}{Def.}[section]
\newtheorem{Thm}[Def]{Theorem}

\newtheorem{Lemma}[Def]{Lemma}

\newcommand{\Proof}{{\em{Proof: }}}
\newcommand{\QED}{\ \hfill $\FBox$ \\[1em]}
\newcommand{\spc}{\;\;\;\;\;\;\;\;\;\;}
\newcommand{\bra}{\mbox{$< \!\!$ \nolinebreak}}
\newcommand{\ket}{\mbox{\nolinebreak $>$}}

\newcommand{\R}{\mbox{\rm I \hspace{-.8 em} R}}
\newcommand{\1}{\mbox{\rm 1 \hspace{-1.05 em} 1}}
\newcommand{\Z}{\mbox{\rm \bf Z}}
\newcommand{\sZ}{\mbox{\rm \bf \scriptsize Z}}

\newcommand{\Tr}{\mbox{Tr\/}}

\newcommand{\FBox}{\rule{2mm}{2.25mm}}

\begin{document}
\include{epsf}

\maketitle

\begin{abstract}
We prove that, in the non-extreme Kerr-Newman black hole geometry, 
the Dirac equation has no normalizable, time-periodic solutions. A 
key tool is Chandrasekhar's separation of the Dirac equation in this 
geometry. A similar non-existence theorem is established in a more 
general class of stationary, axisymmetric metrics in which the Dirac 
equation is known to be separable.
These results indicate that, in contrast with the classical situation 
of massive particle orbits, a quantum mechanical Dirac particle must 
either disappear into the black hole or escape to infinity.
\end{abstract}

\section{Introduction}
\setcounter{equation}{0}

It has recently been proved in \cite{FSY} that the Dirac equation does not
admit normalizable, time-periodic solutions in a non-extreme Reissner-Nordstr\"om
black hole geometry. This result shows that quantization and the introduction
of  spin cause a significant qualitative break-down of the classical situation, where it
is well-known that there exist special choices of initial conditions for the motion of
massive test particles which give rise to closed orbits \cite{C1}. Indeed, the above
theorem implies that the quantum mechanical Dirac wave function
in the gravitational and electromagnetic fields of a static, spherically
symmetric black hole describes a particle which must either disappear into the
black hole or escape to infinity.

It is quite natural to ask whether this result is stable under perturbations of the
background metric; i.e., if the non-existence theorem for normalizable periodic solutions
of the Dirac equation remains true if the background metric and electromagnetic field are
changed in such a way that the spherical symmetry is destroyed. This is precisely the
question that we address in this paper.

We are guided in our choice of a more general background geometry by the uniqueness
theorems of Carter, Israel, and Robinson~\cite{C, CB}, from which we know that the most
general charged black hole equilibrium state is given by the Kerr-Newman solution of the
Einstein-Maxwell equations. Thus, in order to investigate the non-existence of periodic
solutions in the most general black hole geometry, we have to study the Dirac
equation in the Kerr-Newman background. The fact that this study is 
even possible rests on the
remarkable discovery made by Chandrasekhar that the Dirac equation is completely
separable into ordinary differential equations in the Kerr-Newman background geometry,
even though the metric is stationary and axisymmetric
\cite{C2}. In order to state our result, let us first recall that the Kerr-Newman
solution is characterized by three parameters, namely its mass $M$, angular momentum
$aM$ and electric charge $Q$. We prove that the non-existence theorem for
normalizable solutions of the Dirac equation remains true in the case of a rotating
Kerr-Newman black hole provided that the angular momentum per unit mass and the charge
are sufficiently small relative to the total mass of the black hole. Such black holes have
both a Cauchy and an event horizon, and are thus referred to as {\it non-extreme} black
holes. Specifically, we prove:
\begin{Thm}
\label{thm1} In a Kerr-Newman black hole for which $a^2 + Q^2 < M^2$, the Dirac equation
has no normalizable, time-periodic solutions.
\end{Thm}
In Theorem~\ref{thm2}, we give a similar non-existence result for the 
most general stationary, axisymmetric metric in which the Dirac 
equation can be separated by Chandrasekhar's method. The hypotheses
of this theorem are stated as much as possible in geometric terms. While the
solutions one obtains by imposing the Einstein-Maxwell equations on this metric are of
limited physical  interest, this theorem indicates that our non-existence result applies
in a broader context (e.g.\ for a more general energy-momentum tensor).

This paper is organized as follows. In Section~\ref{sec21}, we derive the
separability of the Dirac equation in the Kerr-Newman geometry in a form
slightly different from the one used by Chandrasekhar so as to recover
the equations established in \cite{FSY} in the spherically symmetric limit,
by letting the angular momentum parameter $a$ tend to zero.
In Section~\ref{sec22}, we work out matching conditions for the spinor field
across the horizons. This gives rise
to a weak solution of the Dirac equation in the physical region of the maximal
analytic extension of the Kerr-Newman solution,  which is valid across the
Cauchy and event horizons. We then proceed in Section~\ref{sec23} to establish
the non-existence of time-periodic
solutions. Just as in the spherically symmetric case, the crucial step
consists in exploiting the conservation and positivity of the Dirac current to show that because of the matching
conditions, the only way in which a time-periodic solution of the Dirac equation can be
normalizable is that each term in the Fourier expansion of the spinor field in time and the
angular variable around the axis of symmetry, be identically zero. While the regularity
of the angular dependence of the separable solutions is manifest in the spherically
symmetric Reissner-Nordstr\"om case, this is not so in the axisymmetric case treated in
this paper. The regularity is therefore established in the Appendix. Finally,
we consider in Section~\ref{sec3} the extension of Theorem~\ref{thm1} to more general
stationary axisymmetric metrics.

\section{The Kerr-Newman Black Hole}
\label{sec2}
\setcounter{equation}{0}
\subsection{The Dirac Equation in Boyer-Lindquist Coordinates}
\label{sec21}
Recall that, in Boyer-Lindquist coordinates $(t, r, \vartheta, \varphi)$, the
Kerr-Newman metric takes the form \cite{C}
\begin{eqnarray}
\lefteqn{ ds^2 \;=\; g_{jk}\:dx^j x^k } \nonumber \\
&=& \frac{\Delta}{U} \:(dt \:-\: a \:\sin^2 \vartheta \:d\varphi)^2
\:-\: U \left( \frac{dr^2}{\Delta} + d\vartheta^2 \right) \:-\:
\frac{\sin^2 \vartheta}{U} \:(a \:dt \:-\: (r^2+a^2) \:d\varphi)^2
\;\;, \spc
        \label{eq:0}
\end{eqnarray}
where
\[ U(r, \vartheta) \;=\; r^2 + a^2 \:\cos^2 \vartheta \;\;\;,\spc
\Delta(r) \;=\; r^2 - 2 M r + a^2 + Q^2 \;\;\; . \]
and the electromagnetic potential is of the form
\begin{equation}
A_j \:dx^j \;=\; -\frac{Q \:r}{U} \:(dt \:-\:  a \:\sin^2 \vartheta \:
d\varphi) \;\;\; . \label{eq:1b}
\end{equation}
The metric is singular at the origin $r=0$ and at the zeros of the
function $\Delta$. We shall consider the so-called
{\em{non-extreme case}} $M^2 > a^2 + Q^2$. In this case, $\Delta$ has two
distinct zeros
\[ r_0 \;=\; M \:-\: \sqrt{M^2 - a^2 - Q^2} \spc {\mbox{and}} \spc
r_1 \;=\; M \:+\: \sqrt{M^2 - a^2 - Q^2} \;\;\; . \]
The two radii $r_0$ and $r_1$ correspond to the {\em{Cauchy horizon}} and
the {\em{event horizon}} for the non-extreme Kerr-Newman metric, respectively.

We briefly recall some elementary facts about the Dirac operator in curved
space-time. The Dirac operator $G$ is a differential operator of first order,
\begin{equation}
G \;=\; i G^j(x) \frac{\partial}{\partial x^j} \:+\: B(x) \;\;\;,
        \label{eq:1}
\end{equation}
where $B$ and the Dirac matrices $G^j$ are $(4 \times 4)$-matrices.
The Dirac matrices are related to the Lorentzian metric via the
anti-commutation relations
\begin{equation}
g^{jk}(x) \:\1 \;=\; \frac{1}{2} \:\{G^j(x),\:G^k(x) \} \;\equiv\;
\frac{1}{2} \left(G^j(x) \:G^k(x) \:+\: G^k(x) \:G^j(x) \right)
\;\;\; . \label{eq:2}
\end{equation}
The matrix $B$ is determined by the spinor connection
and the electromagnetic potential through minimal coupling. As such, it
is determined by the Levi-Civita connection of the background Lorentzian
metric (\ref{eq:0}) and the potential (\ref{eq:1b}).
The Dirac matrices are not uniquely determined by the anti-commutation
rules (\ref{eq:2}). The ambiguity in the choice of Dirac matrices adapted
to a given metric is formulated naturally in terms of the spin
and frame bundles \cite{PT}. A convenient
method for calculating the Dirac operator in this bundle formulation
is provided by the Newman-Penrose formalism \cite{C1}.
More generally, it is shown in \cite{F} that all choices of Dirac
matrices satisfying (\ref{eq:2}) yield unitarily equivalent Dirac
operators. Furthermore, in \cite{F}, explicit formulas for the
matrix $B$ in terms of the Dirac matrices $G^j$ are given. In the following, we
attempt to combine the advantages of these different approaches; 
namely, we first choose the Dirac matrices using a Newman-Penrose 
frame, and then construct the matrix $B$ using the explicit formulas in \cite{F}.

We choose the so-called {\em{symmetric frame}} $(l, n, m,
\overline{m})$ of \cite{CM},
\begin{eqnarray*}
l &=& \frac{1}{\sqrt{2 U \:|\Delta|}} \left( (r^2+a^2)
\:\frac{\partial}{\partial t} \:+\: \Delta \:\frac{\partial}{\partial
r} \:+\: a \:\frac{\partial}{\partial \phi} \right) \\
n &=& \frac{\epsilon(\Delta)}{\sqrt{2 U \:|\Delta|}} \left( (r^2+a^2)
\:\frac{\partial}{\partial t} \:-\: \Delta \:\frac{\partial}{\partial
r} \:+\: a \:\frac{\partial}{\partial \phi} \right) \\
m &=& \frac{1}{\sqrt{2 U}} \left( i a \:\sin \vartheta
\:\frac{\partial}{\partial t} \:+\: \frac{\partial}{\partial
\vartheta} \:+\: \frac{i}{\sin \vartheta} \:\frac{\partial}{\partial
\varphi} \right) \\
\overline{m} &=& \frac{1}{\sqrt{2 U}} \left(-i a \:\sin \vartheta
\:\frac{\partial}{\partial t} \:+\: \frac{\partial}{\partial
\vartheta} \:-\: \frac{i}{\sin \vartheta} \:\frac{\partial}{\partial
\varphi} \right) \;\;\;,
\end{eqnarray*}
where $\epsilon$ is the step function $\epsilon(x)=1$ for $x \geq 0$ and
$\epsilon(x)=-1$ otherwise. (Because of its symmetry properties, this 
frame is somewhat more convenient than
the Kinnersley frame used in \cite{C1}. Also, the notation with the step function
allows us to give a unified form of the frame both inside and outside
the horizons.) The symmetric frame is a
Newman-Penrose null frame; i.e.
\[ \bra l, n \ket \;=\; 1 \;\;\;,\spc \bra m, \overline{m} \ket \;=\;
-1 \;\;\; , \]
and all other scalar products between the elements of the frame vanish.
From this complex null frame, we can form a real frame
$(u_a)_{a=0,\ldots,3}$ by setting
\begin{eqnarray*}
u_0 &=& \frac{\epsilon(\Delta)}{\sqrt{2}} \:(l + n) \;\;\;\:, \spc
u_1 \;=\; \frac{1}{\sqrt{2}} \:(l - n) \\
u_2 &=& \frac{1}{\sqrt{2}} \:(m + \overline{m}) \;\;\;, \spc
u_3 \;=\; \frac{1}{\sqrt{2}\: i} \:(m - \overline{m}) \;\;\;.
\end{eqnarray*}
This frame is orthonormal; i.e.
\begin{equation}
g_{jk} \:u^j_a \:u^k_b \;=\; \eta_{ab} \;\;\;,\spc
\eta^{ab} \:u^j_a\:u^k_b \;=\; g^{jk} \;\;\; , \label{eq:3}
\end{equation}
where $\eta_{ab} = \eta^{ab} = {\mbox{diag}}(1,-1,-1,-1)$ is the
Minkowski metric.
We choose the Dirac matrices $\gamma^a$, $a=0,\ldots,3$ of
Minkowski space in the Weyl representation
\[ \gamma^0 \;=\; \left( \begin{array}{cc} 0 & -\1 \\ -\1 & 0
\end{array} \right) \;\;\;,\spc
\vec{\gamma} \;=\; \left( \begin{array}{cc} 0 & -\vec{\sigma} \\
\vec{\sigma} & 0 \end{array} \right) \spc , \]
where $\vec{\sigma}$ are the usual Pauli matrices
\begin{equation}
\sigma^1 \;=\; \left( \begin{array}{cc} 0 & 1 \\ 1 & 0
\end{array} \right) \;\;\;,\spc
\sigma^2 \;=\; \left( \begin{array}{cc} 0 & -i \\ i & 0
\end{array} \right) \;\;\;,\spc
\sigma^3 \;=\; \left( \begin{array}{cc} 1 & 0 \\ 0 & -1
\end{array} \right) \;\;\;.
\label{eq:25a}
\end{equation}
The $\gamma^a$ satisfy the anti-commutation relations
\begin{equation}
\eta^{ab} \;=\; \frac{1}{2} \left\{ \gamma^a, \:\gamma^b \right\}
\;\;\;. \label{eq:4}
\end{equation}
We choose as Dirac matrices $G^j$ associated to the Kerr-Newman metric
the following linear combinations of the $\gamma^a$:
\begin{equation}
G^j(x) \;=\; u^j_a(x) \:\gamma^a
        \label{eq:5}
\end{equation}
As an immediate consequence of (\ref{eq:3}) and (\ref{eq:4}), these
Dirac matrices satisfy the anti-commutation relations
(\ref{eq:2}). Next, we must calculate the matrix $B$ in (\ref{eq:1}).
In \cite{F}, it is shown that the Dirac matrices
induce a spin connection $D$, which has the general form
\begin{eqnarray}
D_j &=& \frac{\partial}{\partial x^j} \:-\: i E_j \:-\: i e A_j
\spc {\mbox{with}}  \\
E_j &=& \frac{i}{2}\: \rho (\partial_j \rho) \:-\: \frac{i}{16}\: \Tr
        (G^m \:\nabla_j G^n) \: G_m G_n \:+\: \frac{i}{8}\: \Tr (\rho G_j \:
        \nabla_m G^m) \:\rho \;\;\; , \label{eq:4b}
\end{eqnarray}
where $\rho = \frac{i}{4!} \epsilon_{jklm} G^j G^k G^l G^m$ is the
pseudoscalar matrix, $\epsilon_{jklm}$ is the Levi-Civita symbol of
curved space-time, and $A_j$ is the electromagnetic potential.
Using the spin connection, the Dirac operator
(\ref{eq:1}) can be written in the alternative form $G=i G^j D_j$.
Thus the matrix $B$ is given by $B=G^j (E_j + e A_j)$. Since, in our
context, the Dirac matrices $G^j$ are linear combinations
of the $\gamma^a$, the matrix $\rho$ is simply the constant
$\rho \equiv \gamma^5 = i \gamma^0 \gamma^1 \gamma^2 \gamma^3$.
As a consequence, the first and last summands in (\ref{eq:4b}) vanish,
and we obtain
\begin{equation}
B \;=\; -\frac{i}{16} \:\Tr (G^m \:\nabla_j G^n) \: G^j G_m G_n \:+\: e
G^j A_j \;\;\; . \label{eq:28a}
\end{equation}
Using Ricci's lemma,
\[ 0 \;=\; 4 \nabla_j g^{mn} \;=\; \nabla_j \Tr (G^m \:G^n) \;=\;
\Tr ((\nabla_j G^m) \:G^n) \:+\: \Tr (G^m \:(\nabla_j G^n)) \;\;\; , \]
and the commutation rules of the Dirac matrices, we can simplify the
trace term as
\begin{eqnarray}
\lefteqn{ \Tr (G^m \:\nabla_j G^n) \: G^j G_m G_n } \nonumber \\
&=& \Tr (G^m \:\nabla_j G^n) \left( \delta^j_m \:G_n \:-\: \delta^j_n
\:G_m \:+\: G^j \:g_{mn} \:+\: i \epsilon^j_{mnp} \:\gamma^5 G^p
\right) \nonumber \\
&=& -8 \:\nabla_j G^j \:+\: i \epsilon^{jmnp} \:\Tr (G_m \:\nabla_j
G_n) \:\gamma^5 G_p \;\;\; . \label{eq:9b}
\end{eqnarray}
Since the Levi-Civita connection is torsion-free, we can replace the
covariant derivative in the last summand in (\ref{eq:9b}) by a partial
derivative. We thus conclude that
\begin{eqnarray*}
B &=& \frac{i}{2 \:\sqrt{|g|}} \:\partial_j (\sqrt{|g|}
u^j_a) \:\gamma^a \\
&& \:-\: \frac{i}{4} \:\epsilon^{jmnp} \:\eta^{ab}\:
u_{am} \:(\partial_j u_{bn})
\:u_{cn}\: \gamma^5 \gamma^c + e A_j u^j_a \:\gamma^a
\end{eqnarray*}
($g$ denotes as usual the determinant of the Lorentzian metric).
This formula for $B$ is particularly convenient, because it only
involves partial derivatives, so that it becomes unnecessary to
compute the Christoffel symbols of the Levi-Civita connection.
Next, we substitute the derived formulas for $G^j$ and $B$ into
(\ref{eq:1}) and obtain for the Dirac operator
\begin{eqnarray}
G &=& \left( \begin{array}{cccc} 0 & 0 & \alpha_+ & \beta_+ \\
0 & 0 & \beta_- & \epsilon(\Delta) \:\alpha_- \\
\epsilon(\Delta) \:\overline{\alpha}_- & -\overline{\beta}_+ & 0 & 0 \\
-\overline{\beta}_- & \overline{\alpha}_+ & 0 & 0
\end{array} \right) \spc {\mbox{with}} \label{eq:16} \\
\beta_\pm &=& \frac{1}{\sqrt{U}} \left( i \frac{\partial}{\partial
\vartheta} \:+\: i \:\frac{\cot \vartheta}{2} \:+\: \frac{a \:\sin
\vartheta}{2 U} \:(r-ia \:\cos \vartheta) \right) \pm
\frac{1}{\sqrt{U}} \left(a \:\sin \vartheta
\:\frac{\partial}{\partial t} \:+\: \frac{1}{\sin \vartheta}
\:\frac{\partial}{\partial \varphi} \right) \nonumber \\
\overline{\beta}_\pm &=& \frac{1}{\sqrt{U}} \left( i \frac{\partial}{\partial
\vartheta} \:+\: i \:\frac{\cot \vartheta}{2} \:-\: \frac{a \:\sin
\vartheta}{2 U} \:(r+ia \:\cos \vartheta) \right) \pm
\frac{1}{\sqrt{U}} \left(a \:\sin \vartheta
\:\frac{\partial}{\partial t} \:+\: \frac{1}{\sin \vartheta}
\:\frac{\partial}{\partial \varphi} \right) \nonumber \\
\alpha_\pm &=& -\frac{\epsilon(\Delta)}
{\sqrt{U \:|\Delta|}} \left( i (r^2+a^2)
\:\frac{\partial}{\partial t} \:+\: i a \:\frac{\partial}{\partial
\varphi} \:+\: e Q r \right) \nonumber \\
&&\hspace*{3.5cm}
\pm \sqrt{\frac{|\Delta|}{U}} \left( i \frac{\partial}{\partial r}
\:+\: i \:\frac{r-M}{2 \Delta} \:+\: \frac{i}{2U} \:(r-ia \cos
\vartheta) \right) \nonumber \\
\overline{\alpha}_\pm &=& -\frac{\epsilon(\Delta)}
{\sqrt{U \:|\Delta|}} \left( i (r^2+a^2)
\:\frac{\partial}{\partial t} \:+\: i a \:\frac{\partial}{\partial
\varphi} \:+\: e Q r \right) \nonumber \\
&&\hspace*{3.5cm} \pm \sqrt{\frac{|\Delta|}{U}} \left( i
\frac{\partial}{\partial r} \:+\: i \:\frac{r-M}{2 \Delta} \:+\:
\frac{i}{2U} \:(r+ia \cos \vartheta) \right) \;\;\;. \nonumber
\end{eqnarray}
The four-component wave function $\Psi$ of a
Dirac particle is a solution of the Dirac equation
\begin{equation}
(G-m) \:\Psi \;=\;
\left( \begin{array}{cccc} -m & 0 & \alpha_+ & \beta_+ \\
0 & -m & \beta_- & \epsilon(\Delta) \:\alpha_- \\
\epsilon(\Delta) \:\overline{\alpha}_- & -\overline{\beta}_+ & -m & 0 \\
-\overline{\beta}_- & \overline{\alpha}_+ & 0 & -m
\end{array} \right) \Psi \;=\; 0 \;\;\;. \label{eq:29a}
\end{equation}

For $a \rightarrow 0$, the Kerr-Newman metric goes over to the
spherically symmetric Reissner-Nordstr\"om metric. Before going on,
we explain how one can recover the Dirac operator of~\cite{FSY} in this
limit: First of all, in \cite{FSY} the Dirac (and not the Weyl)
representation of the Dirac matrices is used. Furthermore, instead of working
with orthonormal frames, the Dirac matrices are constructed in
\cite{FSY} by multiplying the Dirac matrices of Minkowski
space in polar coordinates $\gamma^t$, $\gamma^r$,
$\gamma^\vartheta$, $\gamma^\varphi$ with appropriate scalar functions.
Because of these differences, the operator $G^{a=0}$ obtained from
(\ref{eq:16}) in the limit $a \rightarrow 0$ coincides with the Dirac
operators $G^{\mbox{\scriptsize{in/out}}}$ in \cite{FSY}
only up to a unitary transformation. More precisely, we have, in the
notation of \cite{FSY},
\begin{eqnarray*}
G^{\mbox{\scriptsize{in/out}}} &=& V \:G^{a=0}\:V^{-1} \spc {\mbox{with}} \\
V(\vartheta, \varphi) &=& \frac{1}{\sqrt{2}}
\left( \begin{array}{cc} \1 & -\1 \\ \1 & \1 \end{array} \right) \;
\frac{1}{2} \left(1 + i (\sigma^1 + \sigma^2 + \sigma^3) \right) \:
\exp \!\left(-i\:\frac{\varphi}{2} \:\sigma^1 \right)\:
\exp \!\left(-i\:\frac{\vartheta}{2}\: \sigma^2 \right) \;\;\; .
\end{eqnarray*}
We point out that the unitary transformation $V$ is not continuous, as it
changes sign along the line $\varphi=0$. Since in the limit of flat
Minkowski space, $G^{\mbox{\scriptsize{out}}}$ goes over to the usual
Dirac operator in the Dirac representation, it is natural to assume
(as in~\cite{FSY}) that the Dirac wave function
$\Psi^{\mbox{\scriptsize{in/out}}}$ corresponding to
$G^{\mbox{\scriptsize{in/out}}}$ is continuous (or even smooth away
from the poles $\vartheta=0,\pi$). According to the above transformation
of the Dirac operators, the wave function $\Psi$ in (\ref{eq:29a}) is
given by $\Psi = V^{-1} \Psi^{\mbox{\scriptsize{in/out}}}$, and we conclude
that it must satisfy the boundary conditions
\[ \lim_{\varphi \searrow 0} \Psi(t,r,\vartheta,\varphi) \;=\;
- \lim_{\varphi \nearrow 2 \pi} \Psi(t,r,\vartheta,\varphi) \;\;\;. \]
As a consequence, we must for the eigenvalues of the angular operator
$i \partial_\varphi$ consider half odd integer eigenvalues
(cf.\ (\ref{eq:16a}) below).

It is a remarkable fact that the Dirac equation~(\ref{eq:29a}) can be
completely separated into ordinary differential equations.
This was first shown for the Kerr metric by Chandrasekhar
\cite{C2}, and was later generalized to the Kerr-Newman background
\cite{P, T}. We shall now briefly recall how this is done, since we
will need the explicit form of the corresponding ordinary differential
operators for the proof of Theorem \ref{thm1}. We closely follow
the procedure in \cite{CM}. Let $S(r,\vartheta)$ and
$\Gamma(r, \vartheta)$ be the diagonal matrices\footnote{We
mention for clarity that this transformation of the spinors differs from that
in \cite{CM} by the factor $|\Delta|^{\frac{1}{4}}$ in the definition of $S$.
Our transformation simplifies the radial
Dirac equation; moreover, it will make the form of our matching conditions
easier.}
\begin{eqnarray*}
S &=& |\Delta|^{\frac{1}{4}} \:{\mbox{diag}} \left( (r-ia \:\cos
\vartheta)^{\frac{1}{2}},\: (r-ia \:\cos \vartheta)^{\frac{1}{2}},\:
(r+ia \:\cos \vartheta)^{\frac{1}{2}},\: (r+ia \:\cos \vartheta)^{\frac{1}{2}}
\right) \\
\Gamma &=& -i \:{\mbox{diag}} \left( (r+ia \:\cos
\vartheta),\: -(r+ia \:\cos \vartheta),\:
-(r-ia \:\cos \vartheta),\: (r-ia \:\cos \vartheta) \right) \;\;\; .
\end{eqnarray*}
Then the transformed wave function
\begin{equation}
\hat{\Psi} \;=\; S \:\Psi
\label{eq:21c}
\end{equation}
satisfies the Dirac equation
\begin{equation}
        \Gamma S\:(G-m)\:S^{-1} \:\hat{\Psi} \;=\; 0 \;\;\; .
        \label{eq:11}
\end{equation}
This transformation is useful because the differential operator
(\ref{eq:11}) can be written as a sum of an operator
${\cal{R}}$, which depends only on the radius $r$, and an operator
${\cal{A}}$, which depends only on the angular variable $\vartheta$.
More precisely, an explicit calculation gives
\[ \Gamma S\:(G-m)\:S^{-1} \;=\; {\cal{R}} + {\cal{A}} \]
with
\begin{eqnarray*}
{\cal{R}} &=& \left( \begin{array}{cccc} imr & 0 & \sqrt{|\Delta|}\:
{\cal{D}}_+ & 0 \\
0 & -imr & 0 & \epsilon(\Delta) \:\sqrt{|\Delta|} \:{\cal{D}}_- \\
\epsilon(\Delta) \:\sqrt{|\Delta|} \:{\cal{D}}_- & 0 & -imr & 0 \\
0 & \sqrt{|\Delta|}\:{\cal{D}}_+ & 0 & imr \end{array} \right) \\
{\cal{A}} &=& \left( \begin{array}{cccc}
-am \:\cos \vartheta & 0 & 0 & {\cal{L}}_+ \\
0 & am \:\cos \vartheta & -{\cal{L}}_- & 0 \\
0 & {\cal{L}}_+ & -am\: \cos \vartheta & 0 \\
-{\cal{L}}_- & 0 & 0 & am \:\cos \vartheta \end{array}
\right)\;\;\; , \\
\end{eqnarray*}
where
\begin{eqnarray*}
{\cal{D}}_\pm &=& \frac{\partial}{\partial r} \:\mp\:
\frac{1}{\Delta} \left[ (r^2+a^2) \:\frac{\partial}{\partial t}
\:+\: a \:\frac{\partial}{\partial \varphi} \:-\: i e Q r \right] \\
{\cal{L}}_\pm &=& \frac{\partial}{\partial \vartheta} \:+\: \frac{\cot
\vartheta}{2} \:\mp\: i \left[ a \:\sin \vartheta
\:\frac{\partial}{\partial t} \:+\: \frac{1}{\sin \vartheta}
\:\frac{\partial}{\partial \varphi} \right] \;\;\; .
\end{eqnarray*}
Now for $\hat{\Psi}$ we first employ the ansatz
\begin{equation}
\hat{\Psi}(t,r,\vartheta,\varphi) \;=\; e^{-i \omega t} \:e^{-i (k+
\frac{1}{2}) \varphi} \:
\hat{\Phi}(r, \vartheta)\;\;\;,\spc \omega \in \R, k \in \Z
\label{eq:16a}
\end{equation}
with a function $\hat{\Phi}$, which is composed of radial functions
$X_\pm(r)$ and angular functions $Y_\pm(\vartheta)$ in the form
\begin{equation}
\hat{\Phi}(r,\vartheta) \;=\;
\left( \begin{array}{c} X_-(r) \:Y_-(\vartheta) \\
X_+(r) \:Y_+(\vartheta) \\
X_+(r) \:Y_-(\vartheta) \\
X_-(r) \:Y_+(\vartheta) \end{array} \right) \;\;\; .
\label{eq:16b}
\end{equation}
By substituting (\ref{eq:16a}) and (\ref{eq:16b}) into the transformed
Dirac equation (\ref{eq:11}), we obtain the eigenvalue problems
\begin{equation}
 {\cal{R}} \:\hat{\Psi} \;=\; \lambda \:\hat{\Psi} \;\;\;,\spc
{\cal{A}} \:\hat{\Psi} \;=\; -\lambda \:\hat{\Psi} \;\;\;, \label{eq:17}
\end{equation}
whereby the Dirac equation (\ref{eq:11}) decouples into the system of ODEs
\begin{eqnarray}
\left( \begin{array}{cc} \sqrt{|\Delta|} \:{\cal{D}}_+ & imr - \lambda \\
-imr - \lambda & \epsilon(\Delta) \:\sqrt{|\Delta|}
\:{\cal{D}}_- \end{array} \right)
\left( \begin{array}{c} X_+ \\ X_- \end{array} \right) &=& 0
\label{eq:21a} \\
\left( \begin{array}{cc} {\cal{L}}_+ & -am \cos \vartheta + \lambda \\
am \cos \vartheta + \lambda & -{\cal{L}}_- \end{array} \right)
\left( \begin{array}{c} Y_+ \\ Y_- \end{array} \right) &=& 0 \;\;\; ,
\label{eq:21b}
\end{eqnarray}
where ${\cal{D}}_\pm$ and ${\cal{L}}_\pm$ reduce to the radial and angular
operators
\begin{eqnarray}
{\cal{D}}_\pm &=& \frac{\partial}{\partial r} \:\pm\:
\frac{i}{\Delta} \left[ \omega \:(r^2+a^2)
\:+ \left(k+\frac{1}{2}\right) a  \:+\: e Q r \right] \label{eq:22a} \\
{\cal{L}}_\pm &=& \frac{\partial}{\partial \vartheta} \:+\: \frac{\cot
\vartheta}{2} \:\mp\: \left[ a \omega\:\sin \vartheta
\:+\: \frac{k+\frac{1}{2}}{\sin \vartheta} \right] \;\;\; . \label{eq:22b}
\end{eqnarray}

We mention that in the limit $a \rightarrow 0$, Chandrasekhar's separation
of variables corresponds to the usual separation of the angular dependence
in a spherically symmetric background. The angular operator ${\cal{A}}$
then has the explicit eigenvalues $\lambda=\pm(j+\frac{1}{2})$ with
$j=\frac{1}{2}$, $\frac{3}{2},\ldots$.
The regularity of the eigenfunctions of the angular operator ${\cal A}$
for general $a$ is established in the Appendix.

\subsection{Matching of the Spinors Across the Horizons}
\label{sec22}
Let us consider the Dirac wave function $\Psi = S^{-1} \hat{\Psi}$
with $\hat{\Psi}$ according to the ansatz (\ref{eq:16a}),(\ref{eq:16b}).
Then the Dirac equation  (\ref{eq:29a}) separately describes the wave function
in the three regions $r<r_0$, $r_0<r<r_1$, and $r>r_1$; for clarity, we
denote $\Psi$ in these three regions by $\Psi_I$, $\Psi_M$, and
$\Psi_O$, respectively. Since the ODEs (\ref{eq:21a}),(\ref{eq:21b})
and the transformation $S^{-1}$
are regular for $r \not \in \{ 0, r_0, r_1 \}$, the functions
$\Psi_I$, $\Psi_M$, and $\Psi_O$ are smooth. However, the difficulty is
that the coefficients in the Dirac equation have poles at $r=r_0$
and $r=r_1$. As a consequence, the Dirac
wave function will in general be singular for $r \rightarrow r_{0
\!/\!1}$. Furthermore, is not clear how to treat the
Dirac equation across the horizons.
In this section, we shall derive {\em{matching conditions}}
which relate the wave functions inside and outside each horizon. For the
derivation, we shall first remove the singularities of the metric on the
horizons by transforming to Kerr coordinates.
In these coordinates, we can also arrange that the Dirac operator
is regular. This will allow us to derive a
weak solution of the Dirac equation valid across the Cauchy and event
horizons. In the end, we will transform the derived conditions back to
Boyer-Lindquist coordinates.

First, we must choose coordinate systems where the metric becomes
regular on the horizons. One possibility is to go over to the Kerr coordinates
$(u_+, r, \vartheta, \varphi_+)$ given in infinitesimal form by \cite{C}
\[ du_+ \;=\; dt \:+\: \frac{r^2+a^2}{\Delta} \:dr \;\;\;,\spc
d\varphi_+ \;=\; d\varphi + \frac{a}{\Delta} \:dr \;\;\; . \]
Alternatively, we can choose the coordinates $(u_-, r, \vartheta, \varphi_-)$
with
\[ du_- \;=\; dt \:-\: \frac{r^2+a^2}{\Delta} \:dr \;\;\;,\spc
d\varphi_- \;=\; d\varphi - \frac{a}{\Delta} \:dr \;\;\; . \]
The variables $u_+$ and $u_-$ are the incoming and outgoing null coordinates,
respectively.
Along the lines $u_\pm={\mbox{const}}$, the variables $t$ and $r$ are related
to each other by
\[ dt \;=\; \mp \frac{r^2+a^2}{\Delta} \:dr \;\;\; . \]
By integration, we see that, for $r \rightarrow r_0$, we have
$t \rightarrow \mp \infty$. Similarly, $\lim_{r \rightarrow r_1}
t = \pm \infty$. The fact that $t$ becomes infinite in the limit $r \rightarrow
r_{0\!/\!1}$ means that the Kerr coordinates describe extensions of the
original Boyer-Lindquist space-time, whereby the Cauchy and event horizons
have moved to points at infinity. More precisely, the Cauchy horizon
corresponds to the points $(r=r_0,\:u_+=\infty)$ and $(r=r_0,\:u_-=-\infty)$;
the event horizon is at $(r=r_1,\:u_+=-\infty)$ and $(r=r_1,\:u_-=\infty)$.
In other words, the chart $(u_+, r, \vartheta, \varphi_+)$ extends the
Boyer-Lindquist
space-time across the points $(r=r_0, \:t=-\infty)$ and $(r=r_1, \:t=\infty)$,
whereas the chart $(u_-, r, \vartheta, \varphi_-)$ gives an extension
across $(r=r_0, \:t=\infty)$ and $(r=r_1, \:t=-\infty)$.

We next work out the transformation of the wave
functions from Boyer-Lindquist to Kerr coordinates.
We first consider the chart $(u_+, r, \vartheta, \varphi_+)$.
The transformation of the Dirac equation consists of a transformation of the
space-time coordinates and of the spinors. For clarity, we perform
these transformations in two separate steps. Changing only the
space-time coordinates transforms the Dirac matrices to
\begin{eqnarray*}
G^{u_+} &=& G^t \:\frac{\partial u_+}{\partial t} \:+\:
G^r \:\frac{\partial u_+}{\partial r}
\;=\; -\frac{a \sin \vartheta}{\sqrt{U}} \:\gamma^2 \:+\: \frac{r^2+a^2}
{\sqrt{U \:|\Delta|}} \:(\gamma^0 - \gamma^3) \\
G^r &=& -\sqrt{\frac{|\Delta|}{U}} \left(\Theta(\Delta) \:\gamma^3 \:+\:
\Theta(-\Delta) \:\gamma^0 \right) \\
G^\vartheta &=& -\frac{1}{\sqrt{U}} \:\gamma^1 \\
G^{\varphi_+} &=& G^\varphi \:\frac{\partial \varphi_+}{\partial \varphi}
\:+\: G^r \:\frac{\partial \varphi_+}{\partial r}
\;=\; -\frac{1}{\sin \vartheta \:\sqrt{U}} \:\gamma^2 \:+\: \frac{a}
{\sqrt{U \:|\Delta|}} \:(\gamma^0 - \gamma^3) \;\;\;,
\end{eqnarray*}
where $\Theta$ is the Heaviside function $\Theta(x)=1$ for $x \geq 0$ and
$\Theta(x)=0$ otherwise. The matrices $G^{u_+}$ and $G^{\varphi_+}$ are
singular on the horizons. Therefore we transform the spinors and Dirac
matrices according to
\begin{equation}
\Psi \;\rightarrow\; \tilde{\Psi} \;=\; V(r) \:\Psi \;\;\;,\spc
G^j \;\rightarrow\; \tilde{G}^j \;=\; V(r) \:G^j\: V(r)^{-1} \label{eq:g1}
\end{equation}
with
\begin{equation}
V(r) \;=\; \frac{1}{2} \left(|\Delta|^{-\frac{1}{4}}
\:+\: |\Delta|^\frac{1}{4} \right) \1
\:-\: \frac{1}{2} \left(|\Delta|^{-\frac{1}{4}} \:-\: |\Delta|^\frac{1}{4}
\right) \gamma^0 \gamma^3 \;\;\;.
\label{eq:22c}
\end{equation}
The transformed Dirac matrices are
\begin{eqnarray*}
\tilde{G}^{u_+} &=& -\frac{a \sin \vartheta}{\sqrt{U}} \:\gamma^2
\:+\: \frac{r^2+a^2}{\sqrt{U}} \:(\gamma^0 - \gamma^3) \\
\tilde{G}^r &=&-\frac{1}{2 \sqrt{U}} \:\left[ (1-\Delta) \:\gamma^0 \:+\:
(1+\Delta) \:\gamma^3 \right] \\
\tilde{G}^\vartheta &=& -\frac{1}{\sqrt{U}} \:\gamma^1 \\
\tilde{G}^\varphi_+ &=&-\frac{1}{\sin \vartheta \:\sqrt{U}} \:\gamma^2
\:+\: \frac{\sqrt{a}}{U} \:(\gamma^0 - \gamma^3) \;\;\;.
\end{eqnarray*}
Now the Dirac matrices are regular except at the coordinate singularities
$\vartheta=0, \pi$ and at the origin $r=0$. The anti-commutation relations
(\ref{eq:2}) allow us to check immediately
that the metric is indeed regular across the horizons.
The transformed Dirac operator $\tilde{G}$ can be constructed from the
Dirac matrices $\tilde{G}^j$ with the explicit formulas (\ref{eq:28a}) and
(\ref{eq:1}) (these formulas are valid in the same way with an additional
tilde, because the matrices $\tilde{G}^j$ are again linear combinations
of the Dirac matrices $\gamma^j$ of Minkowski space).
From this, we see that all the coefficients of the Dirac operator $\tilde{G}$
are regular across the horizons. According to the transformation (\ref{eq:g1})
of the wave functions, the Dirac operators $G$ and $\tilde{G}$ are related
to each other by
\[ \tilde{G} \;=\; V\: G \:V^{-1} \;\;\; . \]

Since the operator $\tilde{G}$ is regular across the horizons, we can
now study the Dirac equation on the event and Cauchy horizons.
We denote our original wave functions transformed to Kerr coordinates
by $\tilde{\Psi}_I$, $\tilde{\Psi}_M$, and $\tilde{\Psi}_O$.
They are smooth in the regions $r<r_0$, $r_0<r<r_1$, and $r>r_1$ and
satisfy the Dirac equation there. However, they may have singularities
at $r=r_0$ and $r=r_1$. Let us assume that $\tilde{\Psi} :=
\tilde{\Psi}_I + \tilde{\Psi}_M + \tilde{\Psi}_O$ is a generalized 
solution of the Dirac equation across the horizons.
In order to analyze the behavior of the wave function
near the Cauchy horizon, we write $\tilde{\Psi}$ in a neighborhood of
$r=r_0$ in the form
\[ \tilde{\Psi}(u_+, r, \vartheta, \varphi_+) \;=\;
\Theta(r_0-r) \:\tilde{\Psi}_I(u_+, r, \vartheta, \varphi_+) \:+\:
\Theta(r-r_0) \:\tilde{\Psi}_M(u_+, r, \vartheta, \varphi_+) \]
and substitute into the Dirac equation $(\tilde{G}-m) \tilde{\Psi}=0$.
Since $\tilde{\Psi}$ is a solution of the Dirac equation for $r \neq r_0$, ,
we only get a contribution from the derivative of the Heaviside function, i.e.\
in a formal calculation
\begin{eqnarray*}
0 &=& i \tilde{G}^r \:\delta(r-r_0) \:(\tilde{\Psi}_M(u_+, r, \vartheta,
\varphi_+) - \tilde{\Psi}_I(u_+, r, \vartheta, \varphi_+)) \\
&=& -\frac{i}{2 \:\sqrt{U}} \:\delta(r-r_0) \:
(\gamma^0+\gamma^3)
\left( \tilde{\Psi}_M(u_+, r, \vartheta, \varphi_+)
\:-\: \tilde{\Psi}_I(u_+, r, \vartheta, \varphi_+) \right) \;\;\; .
\end{eqnarray*}
To give this distributional equation a precise meaning, we
multiply the above formal identity
by a test function $\eta(r)$ and integrate,
\begin{equation}
0 \;=\; \int_{r_0-\varepsilon}^{r_0+\varepsilon} \eta(r) \:\delta(r-r_0)\:
(\gamma^0+\gamma^3) \left( \tilde{\Psi}_M(u_+, r, \vartheta, \varphi_+)
\:-\: \tilde{\Psi}_I(u_+, r, \vartheta, \varphi_+) \right) \;\;\; .
\label{eq:i1}
\end{equation}
By choosing a function $\eta(r)$ which goes to zero sufficiently fast
for $r \rightarrow r_0$, we can make sense of this integral, even if
$(\gamma^0+\gamma^3) (\tilde{\Psi}_M-\tilde{\Psi}_I)$
is singular in this limit. For
example, we can choose $\eta$ as $\eta=(1+|(\gamma^0+\gamma^3)
(\tilde{\Psi}_B-\tilde{\Psi}_I)|)^{-1} h$, where $h$ is a smooth function.
It must be kept in mind, however, that we cannot choose $\eta$ independently in
the two regions $r<r_0$ and $r>r_0$, because $\eta$ must be smooth at
$r=r_0$. As a consequence, we cannot conclude from (\ref{eq:i1})
that $\tilde{\Psi}_M$ and $\tilde{\Psi}_I$ must both vanish on the Cauchy
horizon. We only get the weaker condition that they have a similar behavior
near this horizon; namely, it is necessary that the following ``jump 
condition'' holds:
\begin{eqnarray}
\lefteqn{ (\gamma^0+\gamma^3) \:(\tilde{\Psi}_M(u_+,r_0+\varepsilon, \vartheta,
\varphi_+) \:-\: \tilde{\Psi}_I(u_+,r_0-\varepsilon, \vartheta, \varphi_+) }
\nonumber \\
&=& o(1+|(\gamma^0+\gamma^3)\:
\tilde{\Psi}_M(u_+, r_0 + \varepsilon, \vartheta, \varphi_+)|)
\spc {\mbox{as $0<\varepsilon \rightarrow 0$.}}
\label{eq:n1}
\end{eqnarray}
On the event horizon, we obtain in the same way the condition
\begin{eqnarray}
\lefteqn{ (\gamma^0+\gamma^3) \:(\tilde{\Psi}_O(u_+,r_1+\varepsilon, \vartheta,
\varphi_+) \:-\: \tilde{\Psi}_M(u_+,r_1-\varepsilon, \vartheta, \varphi_+) }
\nonumber \\
&=& o(1+|(\gamma^0+\gamma^3)\:
\tilde{\Psi}_M(u_+, r_1 - \varepsilon, \vartheta, \varphi_+)|)
\spc {\mbox{as $0<\varepsilon \rightarrow 0$.}}
\label{eq:n2}
\end{eqnarray}

The constructions we just carried out in the chart $(u_+, r, \vartheta,
\varphi_+)$ can be repeated similarly in the coordinates $(u_-, r,
\vartheta, \varphi_-)$. We list the resulting formulas:
The transformation to the chart $(u_-, r, \vartheta, \varphi_-)$ gives for
the Dirac matrices $G^{u_-}$ and $G^{\varphi_-}$
\begin{eqnarray*}
G^{u_-} &=& -\frac{a \sin \vartheta}{\sqrt{U}} \:\gamma^2 \:+\:
\frac{r^2+a^2} {\sqrt{U \:|\Delta|}} \:\epsilon(\Delta)\:
(\gamma^0 + \gamma^3) \\
G^{\varphi_-} &=& -\frac{1}{\sin \vartheta \:\sqrt{U}} \:\gamma^2
\:+\: \frac{a}{\sqrt{U \:|\Delta|}} \:\epsilon(\Delta)\:
(\gamma^0 + \gamma^3) \;\;\;.
\end{eqnarray*}
To make these matrices regular, we transform the spinors
according to (\ref{eq:g1}) with
\[ V(r) \;=\; \frac{1}{2} \left(|\Delta|^{-\frac{1}{4}}
\:+\: \epsilon(\Delta)\:|\Delta|^\frac{1}{4} \right) \1
\:+\: \frac{1}{2} \left(|\Delta|^{-\frac{1}{4}} \:-\: \epsilon(\Delta)\:
|\Delta|^\frac{1}{4} \right) \gamma^0 \gamma^3 \;\;\;. \]
This gives for the transformed Dirac matrices
\begin{eqnarray*}
\tilde{G}^{u_-} &=& -\frac{a \sin \vartheta}{\sqrt{U}} \:\gamma^2
\:+\: \frac{r^2+a^2}{\sqrt{U}} \:(\gamma^0 + \gamma^3) \\
\tilde{G}^r &=&-\frac{1}{2 \sqrt{U}} \:\left[ (1-\Delta) \:\gamma^0 \:-\:
(1+\Delta) \:\gamma^3 \right] \\
\tilde{G}^\vartheta &=& -\frac{1}{\sqrt{U}} \:\gamma^1 \\
\tilde{G}^\varphi_- &=&-\frac{1}{\sin \vartheta \:\sqrt{U}} \:\gamma^2
\:+\: \frac{a}{\sqrt{U}} \:(\gamma^0 + \gamma^3) \;\;\;.
\end{eqnarray*}
By evaluating the Dirac equation across the horizons in the weak sense,
we obtain the conditions
\begin{eqnarray}
\lefteqn{ (\gamma^0-\gamma^3) \:(\tilde{\Psi}_M(u_-,r_0+\varepsilon,
\vartheta, \varphi_-) \:-\:
\tilde{\Psi}_I(u_-,r_0-\varepsilon, \vartheta, \varphi_-)) }
\nonumber \\
&& \;=\; o(1+|(\gamma^0-\gamma^3) \tilde{\Psi}_M(u_-, r_0 + \varepsilon,
\vartheta, \varphi_-)|) \spc {\mbox{as $\varepsilon \rightarrow 0$.}}
\label{eq:n3} \\
\lefteqn{ (\gamma^0-\gamma^3) \:(\tilde{\Psi}_O(u_-,r_1+\varepsilon,
\vartheta, \varphi_-) \:-\:
\tilde{\Psi}_M(u_-,r_1-\varepsilon, \vartheta, \varphi_-)) }
\nonumber \\
&& \;=\; o(1+|(\gamma^0-\gamma^3) \tilde{\Psi}_M(u_-, r_1 - \varepsilon,
\vartheta, \varphi_-)|) \spc {\mbox{as $\varepsilon \rightarrow 0$.}}
\label{eq:n4}
\end{eqnarray}

It remains to transform the conditions (\ref{eq:n1})--(\ref{eq:n4}) back
to Boyer-Lindquist coordinates.  Since the $t$- and $\varphi$-dependence
of our wave function (\ref{eq:16a}) has the form of a plane wave, we
immediately conclude that the condition (\ref{eq:n1}) is also valid
in Boyer-Lindquist coordinates. We do not again write out that we
consider the limit $0<\varepsilon \rightarrow 0$, but note that
this condition was obtained by extending the Boyer-Lindquist space-time
across the point $(r=r_0, t=-\infty)$,
\begin{eqnarray*}
\lefteqn{ (\gamma^0+\gamma^3) \:(\tilde{\Psi}_M(t,r_0+\varepsilon, \vartheta,
\varphi) \:-\: \tilde{\Psi}_I(t,r_0-\varepsilon, \vartheta, \varphi)) } \\
&=& o(1+|(\gamma^0+\gamma^3)\:
\tilde{\Psi}_M(t, r_0 + \varepsilon, \vartheta, \varphi)|) \spc
{\mbox{across $t=-\infty$.}}
\end{eqnarray*}
We next substitute the transformation (\ref{eq:g1}) and (\ref{eq:22c})
of the spinors. Using the identity
\begin{eqnarray*}
(\gamma^0+\gamma^3) \: V^{-1} &=& (\gamma^0 + \gamma^3)\:
\frac{1}{2} \left( (|\Delta|^{-\frac{1}{4}} + |\Delta|^\frac{1}{4} )
\:+\: (|\Delta|^{-\frac{1}{4}} - |\Delta|^\frac{1}{4}) \:\gamma^0
\gamma^3 \right) \\
&=& |\Delta|^{-\frac{1}{4}} \:(\gamma^0+\gamma^3) \;\;\; ,
\end{eqnarray*}
we obtain the condition
\begin{eqnarray}
\lefteqn{ |\Delta|^{-\frac{1}{4}} \:
(\gamma^0+\gamma^3) \:(\Psi_M(t,r_0+\varepsilon, \vartheta,
\varphi) \:-\: \Psi_I(t,r_0-\varepsilon, \vartheta, \varphi)) } \nonumber \\
&=& o(1+|\Delta|^{-\frac{1}{4}} \:|(\gamma^0+\gamma^3)\:
\Psi_M(t, r_0 + \varepsilon, \vartheta, \varphi)|) \spc {\mbox{across
$t=-\infty$.}}
\label{eq:33b}
\end{eqnarray}
Finally, we consider the transformation (\ref{eq:21c}). It also preserves
the factor $(\gamma^0+\gamma^3)$, because
\begin{eqnarray}
(\gamma^0+\gamma^3) \:S
&=& {\mbox{diag}} \left( (r+ia \:\cos\vartheta)^\frac{1}{2},\:
(r+ia \:\cos \vartheta)^\frac{1}{2},\:
(r-ia \:\cos \vartheta)^\frac{1}{2},\: (r-ia \:\cos \vartheta)^\frac{1}{2}
\right) \nonumber \\
&& \times\: |\Delta|^{\frac{1}{4}} \:(\gamma^0+\gamma^3) \;\;\; .
\label{eq:33a}
\end{eqnarray}
The diagonal matrix in this equation is irrelevant because it is
regular on the horizons. The factors $|\Delta|^{-\frac{1}{4}}$ in
(\ref{eq:33b}) are compensated by the factor $|\Delta|^{\frac{1}{4}}$
in (\ref{eq:33a}), and we end up with the simple condition
\begin{eqnarray}
\lefteqn{(\gamma^0+\gamma^3) \:(\hat{\Psi}_M(t,r_0+\varepsilon, \vartheta,
\varphi) \:-\: \hat{\Psi}_I(t,r_0-\varepsilon, \vartheta, \varphi)) }
\nonumber \\
&& \;=\; o(1+|(\gamma^0+\gamma^3)\:
\hat{\Psi}_M(t, r_0 + \varepsilon, \vartheta, \varphi)|)
\spc {\mbox{across $t=-\infty$.}} \label{eq:m1}
\end{eqnarray}
Similarly, the relations (\ref{eq:n2})--(\ref{eq:n4}) transform into
\begin{eqnarray}
\lefteqn{ (\gamma^0+\gamma^3) \:(\hat{\Psi}_O(t,r_1+\varepsilon, \vartheta,
\varphi) \:-\: \hat{\Psi}_M(t,r_1-\varepsilon, \vartheta, \varphi)) }
\nonumber \\
&& \;=\; o(1+|(\gamma^0+\gamma^3)\:
\hat{\Psi}_M(t, r_1 - \varepsilon, \vartheta, t)|) \spc  {\mbox{across
$t=\infty$}} \label{eq:m2} \\
\lefteqn{ (\gamma^0-\gamma^3) \:(\hat{\Psi}_M(t,r_0+\varepsilon,
\vartheta, \varphi) \:-\:
\hat{\Psi}_I(t,r_0-\varepsilon, \vartheta, \varphi)) }
\nonumber \\
&& \;=\; o(1+|(\gamma^0-\gamma^3) \hat{\Psi}_M(t, r_0 + \varepsilon,
\vartheta, \varphi)|) \spc  {\mbox{across
$t=\infty$}}\label{eq:m3} \\
\lefteqn{ (\gamma^0-\gamma^3) \:(\hat{\Psi}_O(t,r_1+\varepsilon,
\vartheta, \varphi) \:-\:
\hat{\Psi}_M(t,r_1-\varepsilon, \vartheta, \varphi)) }
\nonumber \\
&& \;=\; o(1+|(\gamma^0-\gamma^3) \hat{\Psi}_M(t, r_1 - \varepsilon,
\vartheta, \varphi)|) \spc  {\mbox{across
$t=-\infty$.}}
\label{eq:m4}
\end{eqnarray}
The equations (\ref{eq:m1})--(\ref{eq:m4}) are our matching conditions.

\subsection{Non-Existence of Time-Periodic Solutions}
\label{sec23}
\begin{figure}[tb]
        \centerline{\epsfbox{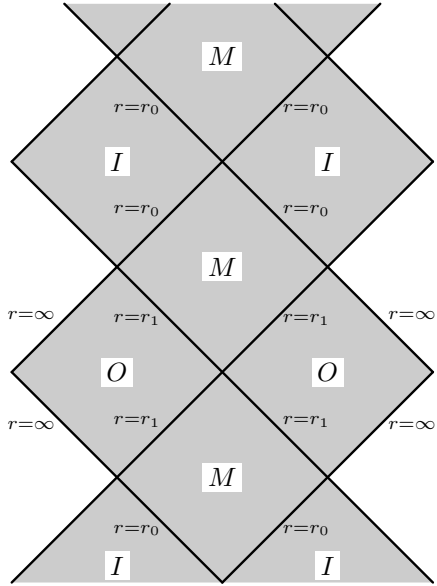}}
        \caption{The conformal structure of the non-extreme
        Kerr-Newman solution}
        \label{fig1}
\end{figure}

Before giving the proof of Theorem \ref{thm1}, we must
specify our assumptions on space-time and on the Dirac wave function in
mathematical terms.
By patching together the Kerr coordinate charts, one obtains the
maximally extended Kerr-Newman space-time \cite{C}. We do not need the
details of the maximal extension here; it suffices to discuss the Penrose
diagram of Figure \ref{fig1}.
Abstractly speaking, the time and
axial symmetry of the Kerr-Newman solution means that the metric admits
two Killing fields. In order to better visualize these symmetries, one can
isometrically map the regions of type $I$, $M$, and $O$ into the regions
$r<r_0$, $r_0<r<r_1$, and $r>r_1$ of the Boyer-Lindquist coordinates,
respectively; then the Killing fields are simply the
vector fields $\partial_t$ and $\partial_\varphi$.
The mappings to Boyer-Lindquist coordinates are unique up to
the isometries of the Boyer-Lindquist space-time (i.e. rotations around the
symmetry axis, time translations, parity transformations, and, in the
case $a=0$, time reversals).
The Kerr coordinates, on the other hand, allow us to
describe three adjacent regions
of type $I$, $M$, and $O$ including the boundaries
between them. When we speak of Boyer-Lindquist or Kerr coordinates in the
following, we implicitly mean that adjacent regions of the maximal
extension are isometrically mapped to these coordinate charts.
The maximal extension may be too general for a truly physical situation
(if one thinks of a black hole which evolved from a
gravitational collapse in the universe, for example, one may want to
consider only one asymptotically flat region). Therefore we assume that our
{\em{physical space-time}} is a given subset of the maximal extension.
We call each region of type $O$ which belongs to
the physical space-time an {\em{asymptotic
end}}. We assume that, in each asymptotic end, a {\em{time direction}}
is given. Thus we can say that each asymptotic end is connected
to two regions of type $M$, one in the future and one in the past.
We assume that the physical wave function can be extended to the
maximal Kerr-Newman space-time; in the regions which do not belong to the
physical space-time, this extension $\Psi$ shall be identically zero.
We want to consider
a {\em{black hole}}, i.e.\ the situation where particles can disappear
into the event horizon, but where no matter can emerge from
the interior of a horizon. Therefore we assume that $\Psi$ is set identically
zero in all regions of type $M$ which are connected
to the asymptotic ends in the past.
The remaining assumptions can be stated most easily in Boyer-Lindquist
coordinates. Since the phase of a wave function $\Psi$ has no physical
significance, when we say that $\Psi$ is {\em{time-periodic with period $T$}}
we mean that there is a real parameter $\Omega$ such that\footnote{We 
remark that the condition of time-periodicity inside the horizon can 
be weakened to local uniform boundedness in $t$. This is proved by 
an ``averaging argument'' identical to the one given 
in~\cite[Appendix~A]{FSY}.}
\[ \Psi(t+T, r, \vartheta, \varphi) \;=\; e^{-i \Omega T} \:
\Psi(t, r, \vartheta, \varphi) \;\;\; . \]
For time-periodic wave functions, we can separate out the time dependence
in a discrete Fourier series. More precisely, we can write the wave function
as a superposition of the form
\begin{equation}
\Psi(t,r,\vartheta, \varphi) \;=\; e^{-i \Omega t}
\sum_{n, k \in \sZ} \;\sum_{\lambda \in \sigma^n_k({\cal{A}})}
e^{-2 \pi i n \:\frac{t}{T}} \:e^{-i (k+\frac{1}{2}) \varphi} \:
\Phi^{\lambda n k}
\label{eq:22}
\end{equation}
with
\[ \Phi^{\lambda n k}(r, \vartheta) \;=\; S^{-1}(r, \vartheta) \:
\hat{\Phi}^{\lambda n k}(r, \vartheta) \;\;\;,\spc
\hat{\Phi}^{\lambda n k}(r, \vartheta) \;=\;
\left( \begin{array}{c} X_-^{\lambda n k} \:Y_-^{\lambda n k} \\
X_+^{\lambda n k} \:Y_+^{\lambda n k} \\
X_+^{\lambda n k} \:Y_-^{\lambda n k} \\
X_-^{\lambda n k} \:Y_+^{\lambda n k} \end{array} \right)  \;\;\; , \]
where the radial and angular functions $X^{\lambda n k}(r)$ and
$Y^{\lambda n k}(\vartheta)$ satisfy the equations (\ref{eq:21a}) and
(\ref{eq:21b}) with
\[ \omega(n) \;=\; \Omega \:+\: \frac{2 \pi}{T} \: n \;\;\; . \]
The index $\lambda$ in (\ref{eq:22}) labels the
eigenvalue of the operator ${\cal{A}}$ in (\ref{eq:17}); the
set $\sigma^n_k({\cal{A}})$
denotes (for fixed $n$ and $k$) all the possible values of $\lambda$.
As shown in the Appendix, the set $\sigma^n_k({\cal{A}})$ is discrete.
Finally, we specify our {\em{normalization condition}}:
The Dirac
wave functions are endowed with a positive scalar product $(.\:|\:.)$.
For this, one chooses a space-like hypersurface ${\cal{H}}$ together with a
normal vector field $\nu$ and considers for two wave functions $\Psi$ and
$\Phi$ the integral
\begin{equation}
(\Psi \:|\: \Phi)_{\cal{H}} \;:=\; \int_{\cal{H}} \overline{\Psi}
G^j \Phi \:\nu_j \:d\mu \;\;\;,
\label{eq:sp}
\end{equation}
where $\overline{\Psi}=\Psi^* \gamma^0$ is the adjoint spinor, and where $d\mu
=\sqrt{g} \:d^3x$ is the invariant measure on ${\cal{H}}$ ($g$ now denotes the
determinant of the induced Riemannian metric).
In a regular space-time, current conservation $\nabla_j \overline{\Psi}
G^j \Phi=0$ implies that the scalar product (\ref{eq:sp}) is independent
of the choice of the hypersurface. The integrand of
$(\Psi | \Psi)_{\cal{H}}$ has the physical interpretation as the
probability density of the Dirac particle. Therefore one usually normalizes
the wave functions in such a way that $(\Psi \:|\: \Psi)=1$.
In our context, the singularities of the metric make
the situation more difficult. Indeed, the normalization integral for a
hypersurface crossing a horizon is problematic since such a hypersurface
will necessarily fail to be space-like on a set of positive measure.
We will therefore only consider the normalization integral in each
asymptotic end away from the event horizon.
More precisely, we choose for given $r_2>r_1$ the one-parameter family 
of hypersurfaces
\[ {\cal{H}}_{t_2} \;=\; \{ (t,r,\vartheta,\varphi) {\mbox{ with }}
t=t_2,\; r>r_2 \} \;\;\;. \]
For a normalized solution $\Psi$ of the Dirac equation, the
integral $(\Psi \:|\: \Psi)_{{\cal{H}}_{t_2}}$ gives the probability of
the particle to be at time $t_2$ in the region outside the ball of the radius
$r_2$ around the origin; this probability must clearly be smaller than one.
Therefore we impose for a normalizable solution the condition that, in
each asymptotic end,
\begin{equation}
 (\Psi \:|\: \Psi)_{{\cal{H}}_{t_2}} \;<\; \infty \spc {\mbox{for all $t_2$.}}
\label{eq:21x}
\end{equation}

We now begin the non-existence proof
by analyzing the wave function in each asymptotic end in Boyer-Lindquist
coordinates. The following
positivity argument shows that each component of the Dirac wave function
must be normalizable: We average the normalization condition (\ref{eq:21x})
one period and use the infinite series (\ref{eq:22}) to obtain
\begin{eqnarray}
\infty &>& \frac{1}{T} \int_t^{t+T} d\tau \;(\Psi \:|\: \Psi)_{{\cal{H}}_{\tau}} \nonumber \\
&=& \sum_{n, n^\prime} \:\sum_{k, k^\prime} \:\sum_{\lambda, \lambda^\prime}
\frac{1}{T} \int_t^{t+T} d \tau \;e^{-2 \pi i(n^\prime - n)
\:\frac{\tau}{T}} \nonumber \\
&& \hspace*{1.5cm} \times
\int_{{\cal{H}}_{\tau}} e^{-i(k^\prime-k) \varphi} \;
\overline{\Phi^{\lambda n k}(r, \vartheta)} \:\Phi^{\lambda^\prime n^\prime
k^\prime}(r, \vartheta) \; d\mu_{\cal{H}} \;\;\; .
\label{eq:e1}
\end{eqnarray}
Since the plane waves in this formula are integrated over a whole period,
we only get a contribution if $k=k^\prime$ and $n=n^\prime$.
As is shown in the Appendix, the $\vartheta$-integration gives zero unless
$\lambda=\lambda^\prime$, and thus the right hand-side of (\ref{eq:e1})
reduces to
\begin{equation}
\sum_{n, k \in \sZ} \:\sum_{\lambda \in \sigma^n_k({\cal{A}})}
\int_{{\cal{H}}_{t, r_2}}
\overline{\Phi^{\lambda n k}(r, \vartheta)} \:\Phi^{\lambda n
k}(r, \vartheta) \; d\mu_{\cal{H}} \;\;\; .
\label{eq:e2}
\end{equation}
Since the scalar product in (\ref{eq:e2}) is positive, we conclude
that the normalization integral must be finite for each $\Phi^{\lambda nk}$,
\[ (\Phi^{\lambda n k} \:|\: \Phi^{\lambda n k})_{{\cal{H}}_{t}} \;<\;
\infty \spc {\mbox{for all $n, k \in \Z$, $\lambda \in
\sigma^n_k({\cal{A}})$, $t \in \R$.}} \]
In this way, we have reduced our problem to the analysis of
static solutions of the Dirac
equation. This simplification is especially useful because it enables us
to work with the matching conditions of the previous section. In the
following, we again use the notation (\ref{eq:16a}),(\ref{eq:16b}).

\begin{Lemma}
\label{lemma1}
The function $|X|^2$ has finite boundary values on the event horizon.
If it is zero at $r=r_1$, then $X$ vanishes identically for $r>r_1$.
\end{Lemma}
{\Proof}
The radial Dirac equation (\ref{eq:21a}) gives for $r>r_1$
\begin{eqnarray}
\sqrt{|\Delta|} \:\frac{d}{dr} |X|^2 &=& \bra \sqrt{|\Delta|} \:
\frac{d}{dr} X,\:X \ket \:+\: \bra  X,\: \sqrt{|\Delta|} \:
\frac{d}{dr} X \ket \nonumber \\
&=& 2\lambda \:{\mbox{Re}}(\overline{X^+} X^-) \:+\: 2mr \:
{\mbox{Im}}(\overline{X^+} X^-) \label{eq:as0}
\end{eqnarray}
and thus
\begin{equation}
-(|\lambda|+mr) \:|X|^2 \;\leq\; \sqrt{|\Delta|} \:\frac{d}{dr} |X|^2
\;\leq\; (|\lambda|+mr) \:|X|^2 \;\;\; . \label{eq:as1}
\end{equation}
In the case that $|X|^2(r)$ has a zero for $r>r_1$, the uniqueness
theorem for solutions
of ODEs yields that $X$ vanishes identically. In the opposite case
$|X|^2(r)>0$ for all $r>r_1$, we divide (\ref{eq:as1}) by
$\sqrt{|\Delta|} \:|X|^2$ and integrate to get
\[ -\int_r^{r^\prime} (|\lambda|+ms) \:|\Delta|^{-\frac{1}{2}} \:ds \;\leq\;
\left. \log |X|^2 \right|_r^{r^\prime} \;\leq\;
\int_r^{r^\prime} (|\lambda|+ms) \:|\Delta|^{-\frac{1}{2}} \:ds \;\;\;,\spc
r_1<r<r^\prime . \]
Using that the singularity of $|\Delta|^{-\frac{1}{2}}$ at $r=r_1$ is
integrable, we conclude that $\log |X|^2$ has a finite limit at $r=r_1$.
\QED
Combining this lemma with our normalization and matching conditions, we
can show now that the Dirac wave function is identically zero outside the
event horizon:
\begin{Lemma}
\label{lemma2}
$\hat{\Psi}_O$ vanishes identically in each asymptotic end.
\end{Lemma}
{\Proof}
According to our black hole assumption, we can apply the matching condition
(\ref{eq:m4}) with $\hat{\Psi}_M \equiv 0$. Expressed in the radial functions
$X$, this implies that
\begin{equation}
\lim_{r_1<r \rightarrow r_1} X^-(r) \;=\; 0 \;\;\; .
\label{eq:as2}
\end{equation}
Using asymptotic flatness and the transformation
\[ \overline{\Psi} \gamma^0 \Psi \;=\; |\Delta|^{-\frac{1}{2}}
\:U^{-\frac{1}{2}} \; \overline{\hat{\Psi}} \gamma^0 \hat{\Psi} \;\;\; , \]
the normalization condition (\ref{eq:21x}) is equivalent to the integral
condition
\begin{equation}
\int_{r_2}^\infty |X|^2 \:dr \;<\; \infty \spc {\mbox{for $r_2>r_1$.}}
\label{eq:as3}
\end{equation}
As an immediate consequence of the radial Dirac equation (\ref{eq:21a}),
\begin{eqnarray}
\frac{d}{dr} \left( |X^+|^2 - |X^-|^2 \right) \;=\; 0 \;\;\;.
\label{eq:as4}
\end{eqnarray}
Thus the function $|X^+|^2 - |X^-|^2$ is a constant; the normalization
condition (\ref{eq:as4}) implies that this constant must be zero,
\[ |X^+|^2 - |X^-|^2 \;\equiv\; 0 \;\;\; . \]
Together with the condition (\ref{eq:as2}), we obtain
that $\lim_{r_1<r \rightarrow r_1} |X|^2 = 0$. Lemma \ref{lemma1} yields that
$X$, and consequently $\hat{\Psi}_O$, must vanish identically.
\QED
We remark that equation (\ref{eq:as4}) can be interpreted physically as the
conservation of the Dirac current in radial direction (notice that
$\overline{\Psi} G^r \Psi = U^{-\frac{1}{2}} \:(|X^+|^2 - |X^-|^2) \:|Y|^2$).

It remains to show that the wave function also vanishes in the interior of
the horizons. For this, we use the matching conditions and
estimates similar to those in Lemma \ref{lemma1}.\\
{\em{Proof of Theorem \ref{thm1}: }}
According to Lemma \ref{lemma2}, $\Psi$ is identically zero in
all regions of type $O$. We first consider a region of type $M$ in
Boyer-Lindquist coordinates. Crossing the event horizon in the past and in
the future brings us to regions of type $O$ where $\hat{\Psi}$ vanishes.
Thus the matching conditions (\ref{eq:m2}) and (\ref{eq:m4}) yield that
\[ \lim_{r_1>r \rightarrow r_1} (\gamma^0 + \gamma^3)
\:\hat{\Psi}_M(t,r,\vartheta,\varphi) \;=\; 0 \;=\;
\lim_{r_1>r \rightarrow r_1} (\gamma^0 - \gamma^3)
\:\hat{\Psi}_M(t,r,\vartheta,\varphi) \;\;\;, \]
and thus
\begin{equation}
\lim_{r_1>r \rightarrow r_1} \:\hat{\Psi}_M(t,r,\vartheta,\varphi) \;=\; 0
\;\;\; . \label{eq:as5}
\end{equation}
The radial Dirac equation (\ref{eq:21a}) in the region $M$ implies that
\[ \sqrt{|\Delta|} \:\frac{d}{dr} |X|^2 \;=\; 0 \]
(this again corresponds to the conservation of the radial Dirac
flux). Thus $|X|^2(r)$ can only go to zero for $r_1>r \rightarrow r_1$ if
it is identically zero. We conclude that $\Psi$ must be identically zero
in all regions of type $M$.

Finally, we consider a region of type $I$ in Boyer-Lindquist coordinates.
We can cross the Cauchy horizon at $t=\infty$ or $t=-\infty$; this brings
us to regions of type $M$, where $\hat{\Psi}_M \equiv 0$. Thus the
matching conditions (\ref{eq:m1}) and (\ref{eq:m3}) imply that
\begin{equation}
\lim_{r_0<r \rightarrow r_0} \hat{\Psi}_I(t,r,\vartheta,\varphi) \;=\; 0
\;\;\; . \label{eq:as6}
\end{equation}
The radial Dirac equation (\ref{eq:21a}) inside the Cauchy horizon is the
same as in the asymptotic end. Thus the estimate (\ref{eq:as1}) again holds,
and we conclude from (\ref{eq:as6}) that $\hat{\Psi}_I$ vanishes
identically.
\QED

\section{The General Separable Case}
\label{sec3}
\setcounter{equation}{0}
Our goal in the present section is to prove an analogue of Theorem \ref{thm1} for the
most general stationary axisymmetric and orthogonally transitive metric in which the
Dirac equation can be separated into ordinary differential equations by Chandrasekhar's
procedure\footnote {We should point out that there exist metrics admitting only one Killing
vector, in which the Dirac equation is completely separable into ordinary differential
equations \cite{FK}. These metrics are not covered by  Chandrasekhar's approach since
the symmetry operators underlying this different type of separability are necessarily
of order $2$ or higher. In contrast, Chandrasekhar's method always gives rise to
symmetry operators of order $1$ since it fits into Miller's theory of factorizable
separable systems \cite{M}.}. An expression for this metric was determined in
\cite{KM}. When the Einstein-Maxwell equations
are imposed, this metric gives rise to all the generalizations of the Kerr-Newman
solution discovered by Carter
\cite{CH}, as well as to a family of exact solutions for which the orbits of the
two-parameter Abelian isometry group are null surfaces \cite{DKM}.

Our aim will be to state as many of the hypotheses of our theorem as
possible in purely geometric terms. We will thus begin by giving a
geometric characterization of the metric which constitutes the starting point of
\cite{DKM}. We will limit ourselves to the case when the orbits of the isometry group
are time-like 2-surfaces, since the procedure is quite similar in the case of
space-like or null orbits.  We wish to point out that the assumption of orthogonal
transitivity is a very natural one to make in a general relativistic context. Indeed, it
is a classical theorem of  Carter and Papapetrou \cite{Pp, Ca} that if the
energy-momentum tensor satisfies some mild invariance conditions, then every stationary
axisymmetric solution of the Einstein equations has the property of admitting
through every point a 2-surface which is orthogonal to the orbit of the isometry group
through that point, i.e.\ the metric is orthogonally transitive.

Let us first recall from \cite{D} that any four-dimensional metric of
Lorentzian signature admitting a two-parameter Abelian group of isometries
acting orthogonally transitively on time-like orbits admits local coordinates
$(u,v,w,x)$ in which
\begin{equation}
 ds^2 \;=\; T^{-2}\left[(L\:du \:+\:M\:dv)^2
\:-\: (N\:du \:+\: P \:dv)^2 \:-\:\frac{dw^2}{W} \:-\:\frac{dx^2}{X} \right]
\;\;, \label{orth}
\end{equation}
where the metric coefficients $L,\;M,\;N,\;P$ and the conformal factor $T^{-1}$ are
functions of $x$ and $w$ only, and where $X=X(x),\;W=W(w)$. 
Furthermore, we have $LP-MN \neq 0$, $T>0$, $W>0$, $X>0$ in order for 
the metric to have the required Lorentzian signature. It is manifest that the
action on the isometry group generated by the Killing vectors $\frac{\partial}{\partial
u}$ and $\frac{\partial}{\partial  v}$ is orthogonally transitive since the
orbits, which are given by the time-like 2-surfaces $x=c_1,\: w=c_2$, admit the
orthogonal 2-surfaces given by $u=c'_{1},\: v=c'_{2} $. Carter \cite{Ca} proved
furthermore that the four-dimensional Lorentzian metrics admitting an Abelian isometry
group acting orthogonally transitively on non-null orbits have the following remarkable
property: there exists in the isotropy subgroup at every point an element of order two
whose differential ${\cal L}$ maps every vector $X$ in the tangent space to the orbit
through that point to its opposite, and maps every vector
$Y$ in the normal space to the orbit to itself. In the local coordinates $(u,v,w,x)$,
this involutive isometry is given by mapping $(u,v,w,x) \rightarrow
(-u,-v,w,x)$. Recall
now that a Newman-Penrose null frame for the metric (\ref{orth}) is said
to be {\it symmetric} \cite{D} if under the involution ${\cal L}$, we have
\begin{equation} {\cal L}(l)=-n\;\;,;\;\; {\cal  L}(n)=-l\;\;,\;\;\;
{\cal L}(m)=-\overline{m}\;\;,\;\;\;{\cal L}(\overline{m})=-m \;\;\;.
\end{equation}
A symmetric null frame $(l, n, m, \overline{m})$ for the metric (\ref{orth})
is given by
\begin{eqnarray*}
l &=& \frac{T}{\sqrt 2} \left( \frac{1}{LP-MN}\left(
P\:\frac{\partial}{\partial u} \:-\: N\:\frac{\partial}{\partial  v} \right)\:+\:
{\sqrt {W}}
\:\frac{\partial}{\partial w} \right) \\
n &=& \frac{T}{\sqrt 2} \left( \frac{1}{LP-MN}\left(  P\:\frac{\partial}{\partial u}
\:-\: N\:\frac{\partial}{\partial  v} \right)\:-\:{\sqrt {W}}
\:\frac{\partial}{\partial w} \right) \\
m &=& \frac{T}{\sqrt 2} \left( \frac{1}{LP-MN}\left(  -M\:\frac{\partial}{\partial u}
\:+\: L\:\frac{\partial}{\partial  v} \right)\:-\: i{\sqrt {X}}
\:\frac{\partial}{\partial x} \right) \\
\overline{m} &=& \frac{T}{\sqrt 2} \left( \frac{1}{LP-MN}\left(
-M\:\frac{\partial}{\partial u}
\:+\: L\:\frac{\partial}{\partial  v} \right)\:+\: i{\sqrt {X}}
\:\frac{\partial}{\partial x} \right)\;\;\;,
\end{eqnarray*}
Besides the metric~(\ref{orth}),  we will also consider the electromagnetic
vector potential given by
\begin{equation}
{\cal A} = A_{j}dx^{j}=T\left( H \:(L\:du
\:+\:M\:dv) +\: K \:(N\:du
\:+\:P\:dv) \right) \;\;\;, \label{pot}
\end{equation}
where $H=H(w)$ and $K=K(x)$. The 1-form (\ref{pot}) is manifestly 
invariant under the action of the isometry group of the metric 
(\ref{orth}). Note that the Maxwell field 2-form $F=d{\cal A}$ is
skew-invariant under the involution
${\cal L}$, so that the electromagnetic energy-momentum tensor is invariant
under ${\cal L}$, in accordance with the hypotheses of the
Carter-Papapetrou theorem~\cite{Ca}.

It was shown in \cite{KM} that a necessary condition for the Dirac equation
$(G-m) \:\Psi =0$ to be separable in the above symmetric null frame for the
metric (\ref{orth}) and in the electromagnetic vector potential (\ref{pot})
is that the metric functions
$L,\: M,\: N$ and
$P$ satisfy the constraints
\begin{eqnarray*}
\frac{\partial}{\partial x}(\frac{L}{LP-MN}) &=& 0\;\;,\;\;\;
\frac{\partial}{\partial x}(\frac{M}{LP-MN}) \;=\; 0 \;\;\;, \\
\frac{\partial}{\partial w}(\frac{N}{LP-MN}) &=& 0\;\;,\;\;\;
\frac{\partial}{\partial w}(\frac{P}{LP-MN}) \;=\; 0\;\;\;.
\end{eqnarray*}
In geometric terms, these conformally invariant conditions are equivalent to
the requirement  that the flows of the null vectors fields
$l$ and $n$ of the symmetric frame be {\it geodesic} and {\it shear-free},
\begin{eqnarray*}
l^{j} \nabla_{j} l^{k} &=& 0\;\;,\;\;\;\;\;\;
2(\nabla_{j}l_{k}+\nabla_{k}l_{j})-(\nabla^{p}l_{p})g_{jk} \;=\; 0 \;\;\;, \\
n^{j}\nabla_{j}n^{k} &=& 0\;\;,\;\;\;
2(\nabla_{j}n_{k}+\nabla_{k}n_{j})-(\nabla^{p}n_{p})g_{jk} \;=\; 0 \;\;\;.
\end{eqnarray*}
It is noteworthy that these conditions are necessary and sufficient for the
St\"ackel separability in the metric (\ref{orth}) of the Hamilton-Jacobi
equation for the null geodesic flow,
\[ g^{jk}\frac{\partial S}{\partial x^j}\frac{\partial S}{\partial x^k}
\;=\; 0\;\;\;. \]
By implementing these separability
conditions and doing some relabeling of the metric functions and the coordinates
we see (see \cite[Section 3]{DKM}) that the metric can be put in the form
\begin{equation}
 ds^2 \;=\; T^{-2}\left[\frac{W}{Z} \:(e_{1}\:du
\:+\:m\:dv)^2
\:-\:\frac{Z}{W} \:{dw^2}  \:-\:
\frac{X}{Z} \:(e_{2}\:du \:+\: p \:dv)^2 \:-\:\frac{Z}{X} \:dx^2 \right]
\;\;, \label{start}
\end{equation}
where $e_{1}$ and $e_{2}$ are constants and where
\[ m\;=\;m(x)\;\;,\;\;\; p\;=\;p(w)\;\;,\;\;\; Z\;=\;e_{1}p-e_{2}m
\;\;\;. \]
The next necessary condition for the Dirac equation to be solvable by
separation of variables in the symmetric frame for the metric
(\ref{orth}) and the electromagnetic
potential (\ref{pot}) is is that the algebraic structure of the Weyl conformal curvature
tensor be of {\it type D} in the Petrov-Penrose classification
\cite{KM}. This conformally invariant condition is necessary and sufficient
for the separability of the massless Dirac or Weyl neutrino equation
$G \Psi=0$ (see \cite[Theorem 2]{KM}).  The type D condition for the metric
(\ref{start}) can be expressed by the condition that the 1-form
$\omega$ given by
\[ \omega \;:=\;(4Z)^{-1}\:(e_{1}m'(x)\:dw\:+\:e_2 \:p'(w) \:dx) \]
be closed,
\begin{equation}
d\omega = 0.
\label{typeD}
\end{equation}
 We therefore have locally
\[ \omega  \;=\; d{\cal B} \]
for some real-valued function ${\cal B}(w,x)$.

Finally, it was shown in \cite[Theorem 3]{KM} that the massive Dirac equation
$(G-m) \:\Psi =0$ is separable in the symmetric frame for the metric (\ref{start}) if
in addition to the conditions stated above, there exist real-valued functions $g(x)$
and
$h(w)$ such that
\begin{equation} Z^{1/2}\: T^{-1}\;\exp(2i{\cal B})\;=\;h(w)+ig(x)
\;\;\;.
\label{sep}
\end{equation}
This is the only separability condition for the Dirac equation which is not
conformally invariant.

We begin by analyzing the type D condition \ref{typeD}. It was proved in
\cite[Section 4]{DKM} that,if this condition is satisfied, then a set of
coordinates $(u,v,w,x)$ for the metric
\ref{start} can be chosen in such a way that the metric functions $p(w),m(x)$,
the constants
$e_1,e_2$ and the function ${\cal B}$ take one of the following four forms:
\begin{description}
\item{ Case $A$:}
\begin{eqnarray*}
e_1 &=& 1 \;=\; e_2 \;\;,\;\;\;
p(w)\;=\;w^2\;\;,\;\;\; m(x) \;=\; -x^2\;\;, \\
{\cal B} &=& {i\over 4}\:\log\left(\frac {w-ix}{w+ix}\right)\;\;\;.
\end{eqnarray*}

\item{ Case $B_{-}$:}
\begin{eqnarray*}
e_1&=&0 \;\;,\;\;\; e_2\;=\;1 \;\;,\;\;\;
p(w)\;=\;2kw\;\;,\;\;\; m(x) \;=\; -x^2-k^2\;\;, \\
{\cal B}&=&{i\over 4}\:\log\left(\frac {k-ix}{k+ix}\right)\;\;\;.
\end{eqnarray*}

\item{ Case $B_{+}$:}
\begin{eqnarray*}
e_1&=&1 \;\;,\;\;\; e_2\;=\;0\;\;,\;\;\; p(w)\;=\;w^2 + l^2
\;\;,\;\;\; m(x)\;=\;-2lx \;\;, \\
{\cal B} &=& {i\over 4}\:\log\left(\frac {w-il}{w+il}\right)\;\;\;.
\end{eqnarray*}

\item{ Case $C^{00}$:}
\begin{eqnarray*}
e_1&=&1\;\;,\;\;\;e_2\;=\;0\;\;,\;\;\; p(w)\;=\;1 \;\;,\;\;\;m(x)\;=\;0
\;\;, \\
{\cal B} &=& -{\pi/4}\;\;\;.
\end{eqnarray*}
\end{description}
We are using the same labelling for the cases given above as the one used in
\cite{CH, DKM}. Each of these cases admits an invariant characterization which we will not
recall here but which is given in \cite {DKM1}. Next, by imposing the
separability condition (\ref{sep}), we conclude that the conformal factor
$T^{-1}$ must be constant in each of the cases $A,B_{-},B_{+}$, and $C^{00}$. This
constant can be normalized to one by a rescaling of the coordinates.

We will restrict our attention for the rest of this section to Case $A$, which can be
thought of as the generic case. Indeed, we will see below that the solutions one obtains
when imposing the Einstein-Maxwell equations in Case $A$ contain the
Kerr-Newman solution as a special case. On the other hand, it is shown in
\cite{CH, DKM} that the solutions  one obtains when imposing the Einstein-Maxwell
equations in the remaining cases
 have isometry groups of dimension $4$ in Cases $B_{-},B_{+}$ and $6$ in Case
$C^{00}$. These solutions notably include the Taub-NUT and Robinson-Bertotti metrics.
Furthermore,
it is shown in the above references that all these solutions can be obtained from
Carter's
$A$ solution by a suitable limiting process.

The metric in Case A is thus given by
\begin{equation}
 ds^2 \;=\; \frac{W(w)}{w^2+x^2} \:(du\:-\:x^2\:dv)^2
\:-\:\frac{w^2+x^2}{W(w)} \:{dw^2}  \:-\:
\frac{X(x)}{w^2+x^2} \:(\:du \:+\: w^2 \:dv)^2 
\:-\:\frac{w^2+x^2}{X(x)} \:dx^2
\;\;. \label{CaseA}
\end{equation}

The Dirac equation for the metric (\ref{CaseA}) can be separated by a procedure analogous
to the one described in Section 2 for the Kerr-Newman metric. It is proved 
in \cite{KM} that the spinor field given by \footnote{ To facilitate the
comparison with the expression given in (\ref{eq:21c}) for the Kerr-Newman case, we have
chosen to include the factor $W^{\frac{1}{4}}$ in the transformation of the spinor
field. Just as in the Kerr-Newman case, this transformation has the effect of producing
a slightly simpler eigenvalue equation in the variable $w$ than the one obtained in
\cite{KM}.}
\begin{equation} {\Psi}(u,v,w,x) \;=\;  e^{-i(\omega u + k
v)} \:(W(w))^\frac{1}{4} \left(
\begin{array}{c}(w-ix)^{1/2}\;X_-(w)
\:Y_-(x)
\\ (w-ix)^{1/2}\;X_+(w)
\:Y_+(x)
\\ (w+ix)^{1/2}\; X_+(w)
\:Y_-(x) \\ (w+ix)^{1/2}\;X_-(w) \:Y_+(x) \end{array} \right)
\;\;\; ,
\label{eq:sepA}
\end{equation}
where $\omega$ and $k$ are constants, will be a solution of the Dirac equation expressed
in the symmetric frame for the metric (\ref{CaseA}) and the Weyl representation of the
Dirac matrices, if and only if the transformed spinor
\begin{equation}
{\Phi}(w,x) \;=\;
\left( \begin{array}{c} X_-(w) \:Y_-(x) \\ X_+(w) \:Y_+(x) \\ X_+(w)
\:Y_-(x) \\ X_-(w) \:Y_+(x) \end{array} \right) \;\;\; .
\label{eq:transfspin}
\end{equation}
satisfies the eigenvalue problems given by
\begin{equation}
 {\cal{W}} \:\Phi \;=\; \lambda \: \Phi \;\;\;,\spc {\cal{X}} \:\Phi
\;=\; -\lambda \:\Phi \;\;\;, \label{eq:eigenA}
\end{equation}
where
\begin{eqnarray*} {\cal{W}} &=& \left( \begin{array}{cccc} im\:w & 0 & \sqrt{W}\:
{{\cal{D}}_{w}}_+ & 0 \\ 0 & -im\:w & 0 & \:\sqrt{W} \:{{\cal{D}}_{w}}_- \\
\sqrt{W} \:{{\cal{D}}_{w}}_- & 0 & -im\:w & 0 \\ 0 &
\sqrt{W}\:{{\cal{D}}_{w}}_+ & 0 & im\:w \end{array} \right) \\ {\cal{X}} &=& \left(
\begin{array}{cccc} -m\:x& 0 & 0 & \sqrt{X}\:{{\cal{L}}_{x}}_+ \\ 0 & m\:x &
-\sqrt{X}\:{{\cal{L}}_{x}}_- & 0 \\ 0 & \sqrt{X}\:{{\cal{L}}_{x}}_+ & -m\:x & 0 \\
-\sqrt{X}\:{{\cal{L}}_{x}}_- & 0 & 0 & m\:x\end{array}
\right)\;\;\; , \\
\end{eqnarray*}
and ${{\cal{D}}_{w}}_\pm$ and ${{\cal{L}}_{x}}_\pm$ are the ordinary
differential operators defined by
\begin{eqnarray*} {{\cal{D}}_{w}}_\pm &=& \frac{\partial}{\partial w} \:\mp\:
\frac{1}{W(w)} \left[ -i\omega w^{2}\:
\:+ik \:-\: i e H(w) \right] \\
{{\cal{L}}_{x}}_\pm &=&
\frac{\partial}{\partial x} \: \:\mp\: i \frac{1}{X(x)}\left[ i\omega x^2 \:-\:
\:ik  -\frac{1}{4}X'(x)-ieK(x)\right]\;\;\; .
\end{eqnarray*}

As a motivation for our generalization of Theorem~\ref{thm1}, we 
recall from \cite{C} how the Kerr-Newman solution arises as a special 
case of the metric (\ref{CaseA}) and vector potential 
(\ref{pot}). For this, we first impose the Einstein-Maxwell equations
\begin{equation}
R_{ij}-\frac{1}{2}R\:g_{ij}-\Lambda\:
g_{ij}=F_{ik}F_{j}^{k}-\frac{1}{4}g_{ij}F_{kl}F^{kl} \;\;\;,
\end{equation}
which determine the remaining functions $X$, $W$, and the electromagnetic
field ${\cal F} = d{\cal A} $. The general solution is given by
\begin{eqnarray}
W &=& \frac{\Lambda}{3}\;w^4 +f_{2}w^2 +f_{1} w +f_{0}+Q^2+P^2
\label{laststep1} \\
X &=& \frac{\Lambda}{3}\;x^4 -f_{2}x^2 +g_{1} x +f_{0} \\
{\cal A} &=& \frac{1}{w^2+x^2}\:\left(Q \:w\:(du\:-\:x^{2}\:dv)
+P\:x\:(du \:+\: w^2 \:dv)\right) \;\;\;,
\label{laststep2}
\end{eqnarray}
where $Q$ and $P$ denote the electric and magnetic monopole moments and
$f_{0},f_{1},f_{2},g_{1}$ are arbitrary parameters. These solutions were first
discovered by Carter \cite{CH}. The Kerr-Newman solution is obtained as a special
case by putting suitable restrictions on the parameters
$\Lambda,f_{0},f_{1},f_{2},g_{1}$ appearing in the metric and by implementing
through the choice of coordinates
$(u,v,w,x)$ the additional hypothesis that the metric is stationary axisymmetric
(as opposed to simply admitting a two-dimensional Abelian isometry group).
This procedure will serve us as a guide in formulating the conditions under
which the analogue of Theorem \ref{thm1} holds for the Dirac equation in the
metric (\ref{CaseA}).

We first re-label the coordinates $(u,v,w,x)$ as $(t,\varphi,r,\mu)$ and rescale
the ignorable coordinates $t$ and $\varphi$ so as to have
$e_1=1,\;e_2=a$, where $a$ is a constant. Note that while $a$ is freely normalizable at
this stage, it will become an essential non-normalizable parameter once our freedom to
scale the coordinates will have been exhausted. The expressions of the metric functions
$p(r)$ and $m(\mu)$ will thus be slightly different in this normalization of the
coordinates. We have
\[ p(r)\;=\;r^2 \;\;,\;\;\; m(\mu)\;=\;-a\mu^2 \;\;\;. \]

Taking the cosmological constant
$\Lambda$ to be zero, the metric functions $W^2(r)$ and $X^2(\mu)$ appearing in
Carter's $A$ solution will take the form
\[ W(r)\;=\;f_{2}\:r^2 -2M\:r+f_{0}\:a^2+Q^2+P^2\;\;,\;\;\;
X(\mu)=-f_{2}\:\mu^2 +g_{1} \:\mu +f_{0}\;\;\;. \]

We now implement the hypothesis of axisymmetry of the metric by letting $\varphi$  be an
angular polar coordinate adapted to the axis of symmetry of the metric and
taking $\mu$ to be proportional to the cosine of the polar angle 
measured from the axis. The range of $\varphi$ will thus be the interval
$(0,2\pi)$ and the range of $\mu$ will be a bounded open interval.
In order for the metric to be of  hyperbolic signature, we
must require that $X(\mu)$ be positive for $\mu$ varying in that bounded interval.  It
follows that the roots of
$X(\mu)$ must be distinct and that $X(\mu)$ must be positive as $\mu$ varies in the
interval bounded by these roots. This gives the constraints
\[ f_{2}\;>\;0\;\;\;,\spc g_{1}^{2}+4f_{2}f_{0}\;>\;0 \;\;\;. \]
Next, we can ensure that the singular behavior of the metric at the roots of
$X(\mu)$ is caused by nothing more than the usual angular coordinate singularity at
the polar axis by choosing the ignorable coordinates and the remaining parameters
appearing in the metric in such a way that the roots of
$m(\mu)$ and
$X(\mu)$ coincide. Thus, we use  our residual freedom to replace $t$ by a constant
linear combination of $t$ and $\varphi$ to add the same arbitrary constant to $p(r)$ and
$m(\mu)$. The roots of $m(\mu)$ will then be located at the endpoints of a symmetric
interval $(-c,c)$. In order for the roots of $X(\mu)$ to also be located at $c$ and
$-c$, the coefficients $g_1,f_2$ and $f_0$ must necessarily satisfy the
relations
\[ g_{1}\;=\;0 \;\;,\;\;\; f_{2}\;>\;0 \;\;,\;\;\;f_{0}\;>\;0 \;\;\;. \]
We can now set $f_{2}=f_{0}=1$ by rescaling $\mu$ and $r$, so that the range of $\mu$
becomes the interval $(-1,1)$. With these normalizations, the parameter $a$ has now
become an essential parameter in the metric.
The metric functions $p,m,W$ and $X$ thus reduce to
\begin{eqnarray*}
p(r) &=& r^2 + a^2 \;\;,\spc\spc\spc\;\;\;\: m(\mu) \;=\; a(1-\mu^2) \;\;, \\
W(r) &=& r^2-2Mr+a^2+Q^2+P^2\;\;,\;\;\;\;X(\mu) \;=\; 1-\mu^2 \;\;\;.
\end{eqnarray*}
We see that by letting $cos\;\mu = \vartheta$, where $-\pi< \vartheta <\pi$, we recover
the Kerr-Newman metric in Boyer-Lindquist coordinates (\ref{eq:0}).

We are now ready to state a theorem which generalizes Theorem 1.1 to the
most general family of stationary axisymmetric metrics in which the Dirac equation is
solvable by separation of variables.  We will consider a normal form for the
metrics (\ref{CaseA}) of Case
$A$ in which the stationary and axisymmetric character of the metric is made manifest
through an appropriate choice of coordinates and by the imposition of suitable
restrictions on the singularities of the metric functions. This normal form is easily
established by a procedure similar to the one we described above for the Kerr-Newman
metric.

\begin{Thm}
\label{thm2}
Consider the stationary, axisymmetric metric and the vector potential given by
\begin{eqnarray}
 ds^2 &=&\frac{W(r)}{r^2+a^{2}\mu^2} \:(dt\:-a\:(1-\mu^2)\:d\varphi)^2
\:-\:\frac{r^2+a^{2}\mu^2}{W(r)} \:{dr^2}  \nonumber \\
&&\:-\:\frac{X(\mu)}{r^2+a^{2}\mu^2} \:(a\:dt \:-\:( r^2 +a^2)\:d\varphi)^2
\:-\:\frac{r^2+a^{2}\mu^2}{X(\mu)} \:d\mu^2  \;\;, \label{axisform} \\
{\cal A}&=&\frac{1}{r^2+a^{2}\mu^2}\left(H(r)\;(dt
\:-a\:(1-\mu^2)\:d\varphi)+K(\mu)\;(a\:dt \:-\:( r^2
+a^2)\:d\varphi)\right) \;\;,\spc
\end{eqnarray}
where $a>0$, $-\infty<t<\infty$, $0 \leq \varphi < 2 \pi$, $H(r) \in 
C^\infty(\R)$, and $K(\mu) \in C^\infty([-1,1])$. Assume that the 
functions $X(\mu) \in C^\infty([-1,1])$ and $W(r) \in C^\infty(\R)$ 
have simple zeros at $\mu = 1, -1$ and $r=r_1,\ldots,r_N$, so that the 
range of the coordinates is $\mu \in (-1,1)$ and $r \in \R \setminus 
\{r_1,\ldots,r_N\}$. Suppose furthermore that the metric is 
asymptotically Minkowskian,
\[ 0 \;<\; \lim_{r \rightarrow \pm \infty} r^{-2} \:W(r) \;<\; 
\infty \;\;\;. \]
Then the Dirac equation $(G-m) \:\Psi =0$ has
no normalizable time-periodic solutions.
\end{Thm}
This theorem can be thought of as an analogue in the axisymmetric context
of the generalization presented in \cite[Remark 4.3]{FSY} of the non-existence
theorem for normalizable time-periodic solutions of the Dirac equation in the
Reissner-Nordstr\"om background. We can likewise argue here that the proof of
Theorem~\ref{thm2} is similar to that of Theorem~\ref{thm1}.
Indeed, we can see from (\ref{eq:eigenA}) that the separated equations obtained
for the metric (\ref{axisform}) are similar in structure to those obtained
in the Kerr-Newman case. Next, we remark that in view of the
assumptions made in Theorem 3.1 to ensure the regularity of the metric at the
symmetry axis, it follows that the maximal analytic extension of the metric
(\ref{axisform}) will have a conformal diagram similar in structure to the one
obtained for the non-extreme Kerr-Newman metric (see Figure \ref{fig1}).
We thus conclude that matching conditions for the spinor
fields across the horizons take a form identical to the equations
(\ref{eq:m1})--(\ref{eq:m4}) which were obtained for the Kerr-Newman case. The key
observation is then that the assumption that we made about the zeros of $W(r)$ will
guarantee that the estimate required to prove the analogue of Lemma 2.1 is valid in
this more general context. Finally, the regularity of the eigenfunctions of the angular
equation is established by the same procedure as in the Kerr-Newman case, by showing
that the angular operator can be viewed as an essentially self-adjoint
elliptic operator on the $2$-sphere $S^2$ with $C^{\infty}$ coefficients.

We conclude by remarking that non-existence theorems similar to
Theorems~\ref{thm1} and~\ref{thm2} likewise hold true in Cases $B_{+},B_{-}$,
and $C^{00}$.

\appendix
\section{Regularity of the Angular Part}
\setcounter{equation}{0}
In this appendix, we will show that all solutions of the angular
equation (\ref{eq:21b}) are regular. More precisely, we will see that 
these functions are of class $C^\infty$ in the open interval 
$0<\vartheta<\pi$ and uniformly bounded on the closed interval $0 
\leq \vartheta \leq \pi$. The method is to reduce
the problem to an elliptic eigenvalue equation on the 2-sphere, where
standard elliptic regularity theory can be applied.

Consider the PDE
\begin{equation}
\left( i \sigma^1 \left( \frac{\partial}{\partial \vartheta} +
\frac{\cot \vartheta}{2} \right) \:+\: i \sigma^2 \:\frac{1}{\sin \vartheta}
\:\frac{\partial}{\partial \varphi} \:+\:a \omega \:\sin \vartheta
\:\sigma^2 \:-\: a m \:\cos \vartheta \:\sigma^3 \right) \alpha
\;=\; \lambda \:\alpha
    \label{eq:A1}
\end{equation}
for a two-component, complex function $\alpha(\vartheta,\varphi)$,
$0<\vartheta<\pi$, $0<\varphi<2 \pi$, with the boundary conditions
\[ \lim_{\varphi \searrow 0} \alpha(\vartheta,\varphi) \;=\;
-\lim_{\varphi \nearrow 2 \pi} \alpha(\vartheta,\varphi) \]
($\sigma^j$ are again the Pauli matrices (\ref{eq:25a})).
With the ansatz
\[ \alpha \;=\; e^{-i(k+\frac{1}{2})\: \varphi} \:\left(
\begin{array}{c} i Y_-(\vartheta) \\Y_+(\vartheta) \end{array} \right)
\;\;\;, \]
the eigenvalue equation (\ref{eq:A1}) simplifies to (\ref{eq:21b}).
Thus it suffices to show regularity for the solutions of (\ref{eq:A1}).

Unfortunately, the coefficients in (\ref{eq:A1}) have singularities
(which are not just removable by a coordinate 
transformation on $S^2$). However, after performing the transformation
\[ \alpha \;\rightarrow\; \tilde{\alpha} = U \alpha \;\;\;{\mbox{with}}\;\;\;
U(\vartheta, \varphi) \;=\
\exp \!\left(-i\:\frac{\varphi}{2} \:\sigma^3 \right)\:
\exp \!\left(-i\:\frac{\vartheta}{2}\: \sigma^2 \right) \;\;\;, \]
we obtain for $\tilde{\alpha}$ the equation $A \tilde{\alpha} =
\lambda \tilde{\alpha}$ with the smooth operator
\[ A \;=\; i \left( \vec{\sigma} \vec{\nabla} - \sigma^r
\frac{\partial}{\partial r} \right) \:+\: a \omega \:\sin \vartheta
\:\sigma\varphi \:-\: a m \:\cos \vartheta\: \sigma^r \;\;\;, \]
where $\sigma^r$ and $\sigma^\varphi$ denote the Pauli matrices in
polar coordinates,
\begin{eqnarray*}
\sigma^r &=& \sin \vartheta \: \cos \varphi \:\sigma^1 \:+\: \sin
\vartheta \:\sin \varphi \: \sigma^2 \:+\: \cos \vartheta \: \sigma^3 \\
\sigma^\varphi &=& -\sin \varphi \:\sigma^1 \:+\:\cos \varphi \:\sigma^2
\;\;\; .
\end{eqnarray*}
Note that the transformation $U$ is not continuous on $S^2$, as it changes
sign along the line $\varphi=0$. Combined with our above boundary conditions
for $\alpha$, we conclude that $\tilde{\alpha}$ is a continuous function
on $S^2$. Hence we can regard the equation $A \tilde{\alpha}=\lambda
\tilde{\alpha}$ as an elliptic eigenvalue equation on the compact domain
$S^2$.
The operator $A$ is essentially self-adjoint on $C^\infty(S^2)^2 \subset
L^2(S^2)^2$. Thus its square $A^2$ is a positive, essentially self-adjoint
operator with smooth coefficients.
Standard elliptic theory yields that $A^2$ has a purely discrete
spectrum with finite-dimensional eigenspaces and smooth
eigenfunctions. Since the eigenvectors of $A$ are obtained by
diagonalizing $A$ on the finite-dimensional eigenspaces of $A^2$, they
are also smooth.\\[1.5em]
{\em{Acknowledgements:}} We would like to thank McGill University, 
Montr{\'e}al, and the Max Planck Institute for Mathematics in the Sciences,
Leipzig, for their hospitality.

\begin{tabular}{lcl}
\\
Felix Finster & $\;\;\;\;$ & Niky Kamran\\
Max Planck Institute for && Department of Math.\ and Statistics \\
Mathematics in the Sciences && McGill University \\
Inselstr.\ 22-26 && Montr{\'e}al, Qu{\'e}bec \\
04103 Leipzig, Germany && Canada H3A 2K6  \\
Felix.Finster@mis.mpg.de && nkamran@math.McGill.CA \\
\\
Joel Smoller & $\;\;$ & Shing-Tung Yau \\
Mathematics Department && Mathematics Department \\
The University of Michigan && Harvard University \\
Ann Arbor, MI 48109, USA && Cambridge, MA 02138, USA \\
smoller@umich.edu && yau@math.harvard.edu
\end{tabular}

\end{document}